\documentclass{article}
\usepackage{jheppub}
\usepackage{amssymb,amsmath,graphicx,cancel,mathtools}
\usepackage{color}
\usepackage[utf8]{inputenc}
\definecolor{darkgreen}{rgb}{0,0.5,0}
\usepackage{subfigure}

\newcommand{\be}{\begin{equation}}
\newcommand{\ee}{\end{equation}}
\newcommand{\beq}{\begin{equation}}
\newcommand{\eeq}{\end{equation}}
\newcommand{\bea}{\begin{eqnarray}}
\newcommand{\eea}{\end{eqnarray}}

\newcommand{\benum}{\begin{enumerate}}
\newcommand{\eenum}{\end{enumerate}}
\newcommand{\bi}{\begin{itemize}}
\newcommand{\ei}{\end{itemize}}
\newcommand{\Mp}{M_{\rm pl}}
\newcommand{\Trh}{T_{\rm RH}}
\newcommand{\Neff}{N_{\mathrm{eff}}}
\newcommand{\Trha}{T_{\mathrm{RH},a}}
\def\p{{ p}}
\def\pp{{\rm p}}
\def\({\left(}
\def\){\right)}
\def\[{\left[}
\def\]{\right]}
\def\nn{\nonumber}
\def\d{{ d}}
\def\bose{\Upsilon}

\makeatletter
\newcommand*\rel@kern[1]{\kern#1\dimexpr\macc@kerna}
\newcommand*\widebar[1]{%
  \begingroup
  \def\mathaccent##1##2{%
    \rel@kern{0.8}%
    \overline{\rel@kern{-0.8}\macc@nucleus\rel@kern{0.2}}%
    \rel@kern{-0.2}%
  }%
  \macc@depth\@ne
  \let\math@bgroup\@empty \let\math@egroup\macc@set@skewchar
  \mathsurround\z@ \frozen@everymath{\mathgroup\macc@group\relax}%
  \macc@set@skewchar\relax
  \let\mathaccentV\macc@nested@a
  \macc@nested@a\relax111{#1}%
  \endgroup
}
\makeatother

\begin{document}

\title{ Chilly Dark Sectors and Asymmetric Reheating}
\author[a]{Peter Adshead,}
\author[b,c]{Yanou Cui,}
\author[a]{and Jessie Shelton}
\affiliation[a]{Department of Physics, University of Illinois at Urbana-Champaign, Urbana, IL 61801, USA }
\affiliation[b]{Perimeter Institute for Theoretical Physics, Waterloo, Ontario N2L 2Y5, Canada }
\affiliation[c]{Maryland Center for Fundamental Physics, University of Maryland, College Park, MD 20742, USA}
\emailAdd{adshead@illinois.edu, ycui@perimeterinstitute.ca, sheltonj@illinois.edu}
\abstract{In a broad class of theories, the relic abundance of dark
  matter is determined by interactions internal to a thermalized dark
  sector, with no direct involvement of the Standard Model (SM).  We
  point out that these theories raise an immediate cosmological
  question: how was the dark sector initially populated in the early
  universe?  Motivated in part by the difficulty of accommodating large
  amounts of entropy carried in dark radiation with cosmic microwave
  background measurements of the effective number of relativistic
  species at recombination, $N_{\mathrm{eff}}$, we aim to establish
  which admissible cosmological histories can populate a thermal dark
  sector that never reaches thermal equilibrium with the SM.  The
  minimal cosmological origin for such a dark sector is {\em
    asymmetric reheating}, when the same mechanism that populates the
  SM in the early universe also populates the dark sector at a lower
  temperature.  Here we demonstrate that the resulting inevitable
  inflaton-mediated scattering between the dark sector and the SM can
  wash out a would-be temperature asymmetry, and establish the regions
  of parameter space where temperature asymmetries can be generated in
  minimal reheating scenarios.  Thus obtaining a temperature asymmetry
  of a given size either restricts possible inflaton masses and
  couplings or necessitates a non-minimal cosmology for one or both
  sectors.
  As a side benefit, we develop techniques for evaluating collision
  terms in the relativistic Boltzmann equation when the full
  dependence on Bose-Einstein or Fermi-Dirac phase space distributions
  must be retained, and present several new results on relativistic
  thermal averages in an appendix.
}

\maketitle


\section{Introduction}
\label{sec:introduction}

Although we have ample gravitational evidence for the existence of
some form of dark matter (DM) constituting 26\% of the energy budget
of our universe \cite{Ade:2015xua}, its detailed nature and properties remain one of the
greatest outstanding mysteries in particle physics and cosmology. The
lack of observational evidence to date of traditional
weakly-interacting massive particle (WIMP) dark matter candidates in
direct detection, indirect detection, and collider searches has helped
to motivate a recent explosion of interest in a much broader range of
dark matter theories that exhibit a wide variety of interesting and
nontraditional signals. In many of these models, the DM relic
abundance is chiefly determined by self-interactions among a set of
fields that live in a thermalized dark sector, with little to no
involvement of the Standard Model (SM).  Such self-interacting thermal
dark sectors can provide novel solutions to long-standing puzzles in
particle physics or astrophysics, can yield novel signals, and in more
generality represent a generic possibility for the physics of the
invisible universe.

Models that invoke a thermalized dark sector immediately raise a
cosmological question: how was this dark sector populated in the early
universe? One minimal answer is to produce the dark sector through a
(small) interaction with the SM, e.g., the
frequently-considered kinetic mixing between a dark gauge group and SM
hypercharge \cite{Holdom:1985ag, Galison:1983pa, Dienes:1996zr}.
Such an interaction between the hidden sector (HS) and the SM is often
required for other reasons, for instance, to enable the deposition of
the dark sector's entropy into the SM plasma prior to Big Bang
Nucleosynthesis (BBN).  A thermal dark sector can be produced when the
coupling between the HS and the SM is sufficiently large to bring the
two sectors into thermal equilibrium at some temperature in the early
universe. This mechanism for populating a dark sector has several
attractive features, and in particular is relatively insensitive to
the as-yet-unknown evolution of the early universe.

On the other hand, once we specify that a dark sector was once in
thermal equilibrium with the SM, we limit the degree to which its
temperature $T_{HS}$ can subsequently differ from the temperature of
the SM plasma, $T_{SM}$. This causes difficulties for many models of
hidden sector dark matter that rely critically on having a large
asymmetry between $T_{HS}$ and $T_{SM}$ in order to prevent hot
relics, such as mirror neutrinos or dark gauge bosons, from
contributing at unacceptable levels to the expansion of the universe
(see, for example, \cite{Kolb:1985bf, Hodges:1993yb,
  Berezhiani:1995am, Feng:2008mu, Ackerman:mha, Kaplan:2009de,
  Fan:2013yva,Foot:2014uba}). From Planck's current constraints on the
effective number of neutrinos \cite{Ade:2015xua}, we can limit the
{\em total} number of degrees of freedom in a hidden sector that was
once thermalized with the SM to be $g_{*S,HS} < \mathcal{O}(10)$, if
it contains hot relics that contribute as neutrinos to the expansion
of the universe during the formation of the CMB.  This constraint
already cannot be met by many of the above models, and will only
become more stringent with time: the projected capabilities of the
next generation of CMB experiments \cite{Wu:2014hta,
  Abazajian:2013oma, Errard:2015cxa} represent an order of magnitude
improvement over Planck's current sensitivity to free-streaming hot
relics.  While this is one of the most frequently-invoked motivations
to consider a decoupled hidden sector, such hidden sectors are a
generic possibility for the origin of dark matter, and can lead to
qualitatively novel signals \cite{Pospelov:2007mp, Feng:2008mu,
  Pappadopulo:2016pkp, Berlin:2016vnh}.

We can then immediately identify two equally minimal mechanisms to
populate a thermal hidden sector that was never in kinetic equilibrium
with the SM.  First, the HS can have a small interaction with the SM
that never enters equilibrium. In this case, the SM plasma will
continually inject energy into the hidden sector through
out-of-equilibrium processes over a range of temperatures from some
initial value $T_{max}$ of order the reheating temperature, until some
final temperature $T_{end}$ when the interaction becomes negligible
\cite{Faraggi:2000pv, Chu:2011be, Bernal:2015xba, Heikinheimo:2016yds}.  Second,
the physical process that populated the SM itself in the early
universe at $T_{max}$ can also simultaneously produce the HS at a
lower temperature; we will refer to this mechanism as ``asymmetric
reheating'' \cite{Berezhiani:1995am,Hodges:1993yb}.

In the first case, the dark sector will undergo non-adiabatic
evolution during the period of continual energy injection.  Depending
on the size of the temperature asymmetry and the relation of $T_{end}$
to various scales in the hidden sector, the entropy injection can
alter the predictions of hidden sector dark matter theories in
interesting ways, e.g., by suppressing indirect detection signals (see
also \cite{Cheung:2010gj}).

The second option, asymmetric reheating, requires that some physics
beyond the SM couple to both the SM and the HS.  This implies the
inevitable existence of reactions that transfer energy between the two
sectors, and thus can, if strong enough, erase a would-be temperature
asymmetry.  In this paper, we study in detail the energy transfer
between two sectors in minimal single-field models of reheating, and
identify the regions where potentially resonant inflaton or reheaton
exchange efficiently equilibrates the two sectors.  Along the way we
obtain new results on relativistic thermal averages for particles
obeying Bose-Einstein and Fermi-Dirac distributions.  Thus we will
demonstrate that a sufficiently large temperature asymmetry between
two sectors either (1) restricts the possible combinations of
reheating temperatures, inflaton masses, and inflaton coupling
structures; (2) requires a non-minimal mechanism for reheating,
e.g. with multiple fields \cite{Berezhiani:1995am}; or (3) requires an
alternative cosmological history for one or both sectors, e.g. with
late-time asymmetric entropy release into the SM alone
\cite{Randall:2015xza}.

Because reheating occurs on subhorizon scales that subsequently
undergo non-linear evolution, in simple inflationary scenarios the
reheating epoch itself generically leaves no direct imprint on scales
relevant for cosmology, such as could be observed in the CMB or the
distribution of galaxies.\footnote{More exotic scenarios such as
  modulated reheating \cite{Dvali:2003em} or multifield dynamics
  \cite{Bond:2009xx} can be invoked to imprint large-scale density
  fluctuations.} Thus information about the reheating phase of the
universe is generically very difficult to obtain directly.  Aside from
model-dependent effects such as the production of gravitational waves,
magnetic fields, and primordial black holes (see, for example
\cite{Easther:2006gt, Kronberg:1993vk, Khlopov:1985jw, Amin:2014eta}
and references therein), the primary impact of the reheating phase on
cosmology enters through the unknown evolution of the scale factor and
Hubble rate during the entire period that connects the end of
inflation to the period of radiation domination. This epoch of
expansion changes how physical length scales in the universe today are
related to length scales during inflation, and can thereby alter the
precise location on the inflationary potential where the observable
fluctuations in the CMB were produced. Therefore, given a specific
model for inflation which makes a specific prediction for the spectrum
of fluctuations, indirect constraints on reheating can be placed by
constraining the expansion history during the reheating period
\cite{Martin:2010kz, Adshead:2010mc, Martin:2014nya, Dai:2014jja,
  Cook:2015vqa, Domcke:2015iaa}. However, caution is necessary when
translating these
into constraints on microphysical particle physics theories, as the
detailed properties of the processes responsible for reheating can
significantly alter its duration \cite{Dolgov:1989us, Traschen:1990sw,
  Kofman:1994rk, Shtanov:1994ce, Kofman:1997yn, Chung:1998rq,
  Giudice:2000ex, Kolb:2003ke, Drewes:2015coa}. By contrast, the
approach we take here is to remain as independent of specific
inflationary models as possible, and our results are largely
insensitive to the details of the potential probed during inflation,
or the matter content of the SM and hidden sectors. Connecting the
physics of the reheating epoch to admissible cosmologies of dark
matter theories helps expand future avenues to further pin down the
properties of this phase of our universe's evolution.

We begin by remarking on the currently forecast sensitivity to $\Delta
N_{\mathrm{eff}}$ from future CMB experiments, translating the
potential sensitivity into projected constraints on the field content
in once-thermalized hidden sectors that contain a free-streaming hot
relic in section \ref{sec:deltaneff}.  In section \ref{sec:setup} we
discuss single-field reheating and describe the setup for our
calculations in section \ref{sec:examplesReh}, which maps out the
regions where asymmetric reheating can or cannot yield a temperature
asymmetry for a variety of different minimal models.  Section
\ref{sec:conclusions} contains our conclusions.  New analytic and
numerical results on relativistic thermal averages are collected in
the appendices \ref{App:Exfer} and \ref{App:specifics}.  In appendix
\ref{sec:preheating} we present a brief overview of preheating.  We
work in units where $\hbar = c = k_{B} = 1$, but we explicitly retain
the reduced Planck mass $M^2_{\rm Pl} = 1/(8\pi G)$.

\section{Future CMB sensitivity to dark radiation in thermal hidden sectors}
\label{sec:deltaneff}

A long-standing motivation for asymmetric reheating has been the need
to prevent stable dark radiation in a dark sector from contributing at
unacceptable levels to observables sensitive to the expansion of the
universe at early times.  The most important such observables are the
light element abundances from BBN, which test the number of
relativistic species at temperatures $T\sim
\mathcal{O}(\mathrm{MeV})$, and the CMB, which is sensitive to the
number of relativistic species at temperatures $T\sim
\mathcal{O}(\mathrm{eV})$.  While the need to reconcile the
existence of dark radiation with existing BBN and CMB measurements is
hardly the only motivation to consider asymmetric reheating, it arises
in a wide variety of models, and is connected to some of the most
direct experimental signatures in several models of hidden sector dark
matter.  For this reason it is worth mentioning that the anticipated
sensitivities of future CMB experiments, in particular those currently
being discussed for CMB Stage-IV, improve on Planck's current
sensitivity to dark radiation by an order of magnitude. This
improvement has profound implications for dark sectors with stable
relativistic relics that were ever in thermal equilibrium with the SM,
as we briefly review here.

The sensitivity of CMB measurements to the number of effective
relativistic degrees of freedom at recombination is conventionally
quoted in terms of the effective number of neutrinos, $\Neff$.
Additional relativistic degrees of freedom contribute a shift in the
effective number of neutrinos by an amount
\begin{equation} \label{eqn:deltaneff}
\Delta N_{\mathrm{eff}} = 2.2   \,g_{*,HS}^{IR}  \left(\frac{T^{IR}_{HS}}{T^{IR}_{SM}}\right) ^ 4,
\end{equation} 
where $g_{*S,HS}, g_{*,HS}$ are defined (in analogy to the usual
definitions of $g_{*S}, g_*$ in the SM) within the dark sector with
respect to the temperature of the dark plasma $T_{HS}$,
\bea
g_{*S,HS} &=& \sum_{i\in\mathrm{bosons}} g_i
\left(\frac{T_i}{T_{HS}}\right)^3 +\frac{7}{8} \sum_{j\in\mathrm{fermions}} g_j \left(\frac{T_j}{T_{HS}}\right)^3\\
g_{*, HS}&=&\sum_{i\in\mathrm{bosons}} g_i
\left(\frac{T_i}{T_{HS}}\right)^4 +\frac{7}{8} \sum_{j\in\mathrm{fermions}} g_j\left(\frac{T_j}{T_{HS}}\right)^4.
\eea
For simplicity we assume a single dark radiation species, so, denoting
by $IR$ the value pertaining during the formation of the CMB,
$g_{*,HS}^{IR} =g_{*S,HS}^{IR} $, while $g_{*S,SM} ^
{IR}=3.9$.

Current forecasts for the capabilities of next-generation CMB
experiments, in particular CMB Stage-IV, anticipate an exciting level
of sensitivity to $\Delta N_{\mathrm{eff}}$, with a projected 68\% CL
uncertainty in the range $\sigma (N_{\mathrm{eff}}) \sim 0.015-0.03$
\cite{Wu:2014hta, Abazajian:2013oma, Errard:2015cxa}.  To highlight
the potential impact on models of hidden sector dark matter, we
briefly illustrate the implications of a SM-like measurement of
$\Neff$ at this level of precision for the total field
content of simple hidden sectors.

Specifically, we are interested in internally-thermalized hidden
sectors that contain stable dark state(s) that may include dark
radiation as well as dark matter.  When the branching fractions of
hidden sector states to SM states are negligible, all (or almost all)
of the entropy contained in a hidden sector is ultimately carried by
the lightest stable dark state.  In the generic situation where this
lightest stable dark state is a hot relic that is still relativistic
during the formation of the CMB, the total entropy in the hidden
sector governs the magnitude of the dark radiation's contribution
to $\Neff$.  If such a dark sector was ever in thermal equilibrium
with the SM at some point in its cosmic history, then the temperature
at thermal decoupling $T_D$ sets a lower bound on the amount of
entropy in the hidden sector post-decoupling, and therefore on the
dark radiation's ensuing contribution to $\Neff$.

\begin{figure}[t]
\begin{center}\includegraphics[width=0.65\textwidth]{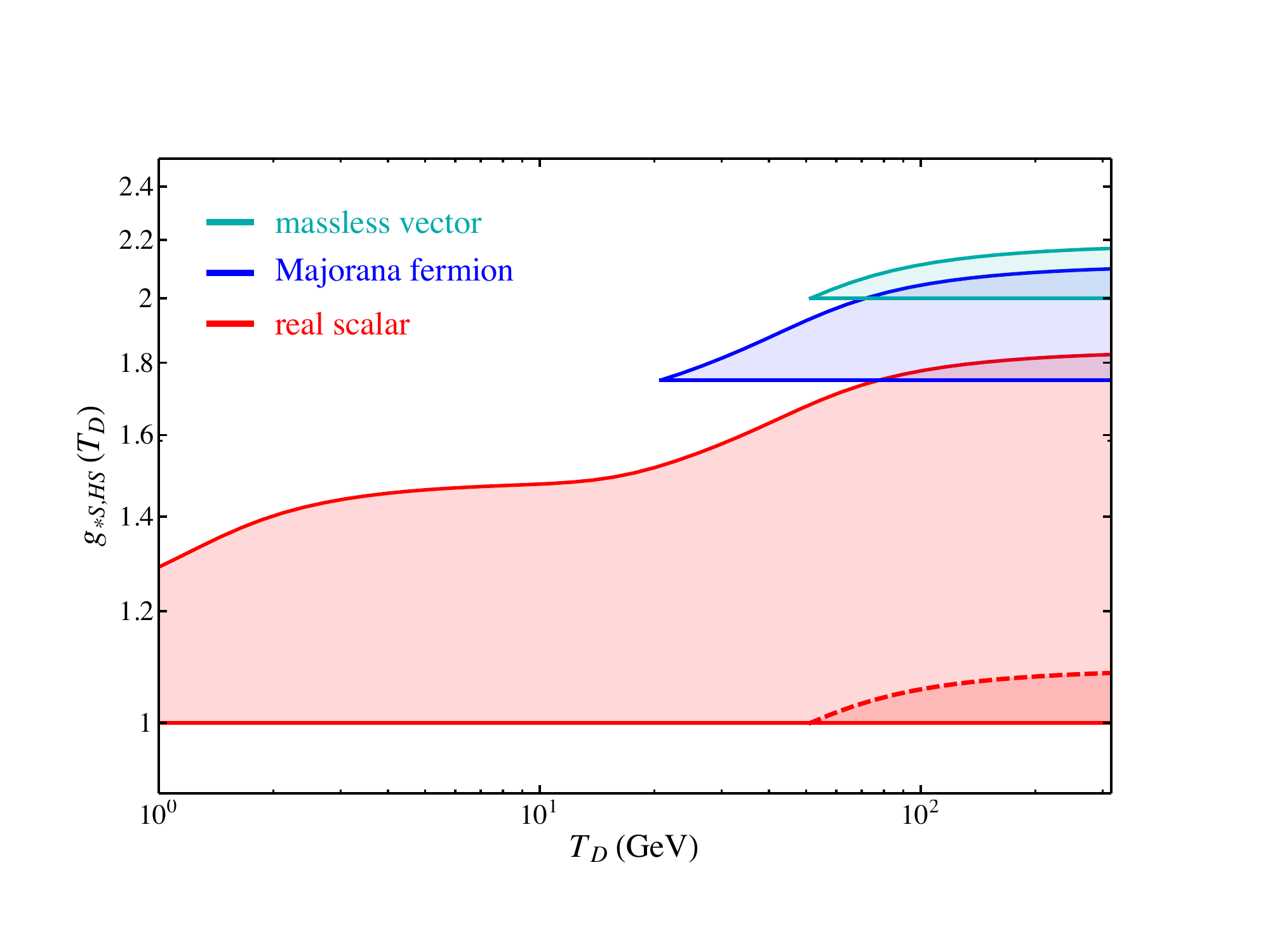}
  \caption{Projected values of $g_{S,*HS} (T_D)$
    allowed at $2\sigma$ by future CMB experiments such as CMB Stage-IV in a hidden sector that contains
    free-streaming dark radiation, as a function of the HS-SM kinetic
    decoupling temperature $T_D$.  Allowed regions are shaded, using
    $\sigma(\Neff) = 0.03$ (solid boundary) and $\sigma(\Neff)=0.015$
    (dotted boundary), for real scalar (red), Majorana fermion (blue),
    or massless vector boson (cyan) dark radiation.}
\label{fig:gHS}
\end{center}
\end{figure}

For simplicity, we will take the dark radiation to be free-streaming,
and work in the limit of instantaneous decoupling.
Under these
assumptions, a constraint on $\Neff$ immediately translates into a
constraint on the total effective number of degrees of
freedom in a hidden sector that was last in equilibrium with the SM at
temperature $T_D$.

In figure \ref{fig:gHS}, we show the projected values of $g_{*S,HS}(T_D)$
that would be consistent at $2\sigma$ with a future measurement of
$\Neff$ at its SM value in CMB Stage-IV, for $\sigma(\Neff)=0.015$ and
$\sigma(\Neff)=0.03$, and for three different species of dark
radiation.
With $\sigma(\Neff)=0.015$, CMB Stage-IV will be able to exclude a single
massless vector or fermionic dark radiation species at $2\sigma$, but
there will remain a small window for real scalars that decouple from
the SM at $T_D > 51$ GeV.  With $\sigma(\Neff)=0.03$, there is some
allowed parameter space remaining for single massless vector or
fermionic dark radiation species at sufficiently large $T_D$, while
real scalars decoupling from the SM at any temperature above the
chiral phase transition are allowed.  Note that none of these
scenarios allow more than one hidden sector species to be relativistic
at $T_D$, which would place stringent constraints on the allowed
masses of possible non-relativistic relics---i.e., dark matter---in
the hidden sector.
These results assume free-streaming dark radiation and
instantaneous decoupling, and are thus are only illustrative.  In
specific models of once-thermalized dark sectors, model-dependent
effects such as the distortion of the phase space distribution
function of the dark radiation during decoupling \cite{Brust:2013xpv}
or the interactions of dark radiation with itself and
with dark matter \cite{Buen-Abad:2015ova, Chacko:2015noa,
  Baumann:2015rya} 
  can often contribute
corrections that will become increasingly important given the
unprecedented sensitivity being discussed for CMB-IV and related
experiments.

One possible way to relax this constraint on the field content of such
hidden sectors is to add new degrees of freedom coupling to the SM in
the ultraviolet.  Adding the field content of the Minimal Supersymmetric Standard Model (MSSM) below $T_D$
relaxes the forecast constraint sufficiently to allow, e.g., a
tiny hidden sector with $g_{*S,HS}^{D} < 3.004$ and $g_{*S,HS}^{IR} =
1$, taking $\sigma(\Neff) = 0.015$.
While this hidden sector can now in principle allow for multiple
species, it cannot accommodate an entire superfield, suggesting that
the scale of supersymmetry breaking must be higher in the dark sector
than in the SM.

A far more flexible way to relax the constraints on dark radiation is
simply not to let the hidden sector ever thermalize with the SM.  If
$T_{HS}$ is sufficiently small compared to $T_{SM}$, models with dark
radiation (such as, for example, \cite{Feng:2008mu, Ackerman:mha,
  Kaplan:2009de, Fan:2013yva}) can be made compatible with even
next-generation measurements of $\Neff$ and related observables.
Given this strong motivation to consider chilly dark sectors, we now
turn to discussing how they can be obtained during the reheating
epoch, and the non-trivial demands that are therefore placed on the
possible reheating histories of our universe.

\section{Perturbative reheating}
\label{sec:setup}

In this section, we briefly review post-inflationary reheating
\cite{Abbott:1982hn,Albrecht:1982mp, Dolgov:1982th, Traschen:1990sw,
  Kofman:1994rk, Shtanov:1994ce, Kofman:1997yn, Chung:1998rq,
  Giudice:2000ex} and outline the main assumptions defining the
minimal model that we adopt in this paper.

In the simplest picture of reheating, dubbed the elementary theory of
reheating \cite{Dolgov:1982th, Albrecht:1982mp, Abbott:1982hn,
  Kofman:1994rk}, reheating occurs through the perturbative decays of
the inflaton as it oscillates about the minimum of its potential.
This process occurs on a time scale set by the inflaton width, $t \sim
1/\Gamma_\phi$. During this time, the energy stored in the inflaton
field dominates the energy density of the universe, resulting in a
matter-dominated expansion. The decays of the inflaton drain the
energy from the condensate and damp the oscillations, leaving a
relativistic bath of daughter particles. Reheating thus ignites the
subsequent radiation-dominated era as relativistic particles are
produced and subsequently thermalize.  This process can be described
by the following Boltzmann equations for the energy densities of the
inflaton and the radiation bath,
\begin{eqnarray}
\label{eq:rhclassic1}
\frac{d\rho_\phi}{dt} + 3 H \rho_\phi  &=& - \Gamma_\phi \rho_\phi,\\
\label{eq:rhclassic2}
\frac{d\rho_R}{dt}+4 H\rho_R &=& \Gamma_\phi\rho_\phi.
\end{eqnarray}
The energy density stored in radiation rapidly increases from zero to
a maximum value and then declines as $T\sim a^{-3/8}$, where $T$ can
always be formally defined through
\begin{align}
\label{eqn:Trad}
T = & \(\frac{30 \,\rho_{R}}{\pi^2 g_*}\)^{1/4},
\end{align}
even before the radiation bath has attained internal thermal
equilibrium.  Here $g_*$ is the number of (effective) degrees of
freedom in the radiation bath.  The reheat temperature $\Trh$ is
defined as the temperature where the radiation comes to dominate over
the inflaton, at which point the universe enters a standard
radiation-dominated phase.  A simple estimate of $\Trh$ can be
obtained by comparing the inflaton decay width $\Gamma_\phi$ to the
Hubble rate. When the decay width is comparable to the Hubble rate,
$\Gamma_\phi\sim H$, the decay process proceeds efficiently.  In the
radiation-dominated era starting from $\Trh$, the Hubble parameter is
given by
\be
H=\left[ \frac{\pi^2}{90} g_* (T)  \right]^{1/2}\frac{T^2}{\Mp},
\ee
which gives the classic expression for $\Trh$ as a function of the
inflaton width,
\begin{align}
\label{eqn:TRH1}
\Trh \approx  \(\frac{90}{\pi^2 g_{\rm RH}}\)^{1/4} \sqrt{\Gamma_{\phi}M_{\rm Pl}}. 
\end{align}
While the maximum value attained by the radiation energy density
depends on the energy scale at the end of inflation, the reheat
temperature does not, and neither quantity is directly dependent on
the inflaton mass.

This classic picture of reheating has made some reasonable but
simplifying assumptions.  In particular,
\bi

\item {\it Reheating is driven by the oscillations and decay of a
    single field $\phi$.}  This is the minimal possibility, and an
  assumption we will maintain throughout this paper. For simplicity,
  we will refer to this field as the inflaton, although it could be a
  separate ``reheaton'' field.\footnote{See e.g.~\cite{Reece:2015lch}
    for a recent discussion of asymmetric reheating from a modulus
    ``reheaton''.}

\item {\it The region of the potential seen by the inflaton during its
    oscillations is purely quadratic.}  Departures from pure quadratic
  behavior will alter the effective equation of state of the
  oscillating inflaton field, which can be parameterized by a
  field-dependent $w(\phi)$ \cite{Turner:1983he}.  A non-zero
  $w(\phi)$ alters the expansion rate of the universe during
  reheating, which will in turn affect the resulting $\Trh$ for fixed
  inflaton mass and width.
  
\item {\it The dissipation of energy from the inflaton condensate
    occurs through perturbative two-body decays.}  Depending on the
  form of the interaction between the inflaton and its daughter fields
  as well as the details of the interaction between the daughter
  fields themselves, this assumption may fail during a large fraction
  of the reheating process.  In particular, there may be significant
  particle production through a non-perturbative ``preheating''
  process involving parametric resonance if the couplings between the
  inflaton and matter are sufficiently strong \cite{Traschen:1990sw,
    Kofman:1994rk,Kofman:1997yn}.  Preheating can result in a far more
  efficient transfer of energy from the inflaton condensate to the
  radiation bath than the tree-level width implies, and will
  generically increase the ultimate $\Trh$ for fixed inflaton mass and
  width. Other possible effects such as thermal blocking and Landau
  damping can arise when we relax the additional
  implicit assumption that collective effects in the radiation bath
  can be largely neglected \cite{Kolb:2003ke, Yokoyama:2005dv,
    Drewes:2010pf, Drewes:2013iaa}.
    
\ei
For our studies in this paper, which serves as the first general study
of multi-sector reheating, we will adopt the following reasonable but
simplifying assumptions as a starting point to describe the radiation bath:
\begin{itemize}

\item {\it All states of interest in the radiation sector have
    effective masses $m\ll M_\phi, T_{RH}$.}  In other words, we
  consider radiation baths composed of particles with masses small
  compared to other mass scales in the problem. This covers the
  majority of the parameter space and allows for simpler analytic
  approximations to various quantities we compute. While the
  production of heavy particles ($m\gtrsim T_{RH}$) during the
  reheating process can be of interest for theories of dark matter and
  baryogenesis (for example, \cite{Chung:1998bt,Chung:1998rq,
    Giudice:1999fb, Giudice:2000ex, Allahverdi:2002pu,Dev:2013yza}), this
  possibility is tangential to our study.  We leave the careful study
  of the effects of thermal masses to future work.

\item {\it Radiation sectors have attained internal thermal
    equilibrium by $T_{RH}$.}  Attaining an equilibrium phase space
  distribution requires number-changing as well as momentum-changing
  interactions to be efficient \cite{Enqvist:1990dp, Davidson:2000er}.
  The timescale for the thermalization of a radiation sector is
  necessarily dependent on the strength and structure of the
  interactions within that sector, and in some cases can be longer
  than $\Gamma_\phi^{-1}$.  For simplicity of presentation, we will
  assume near-instantaneous thermalization in writing reaction rates,
  but all of our quantitative conclusions use these reaction rates
  evaluated at $T_{RH}$ and later, where it will be often be a good
  assumption that local thermal equilibrium has been attained within
  each sector.  In some bosonic cases that we will see below,
  reheating can be very quick, which can render this assumption
  questionable.  However, in these cases, using a thermal distribution
  to describe the radiation is {\it conservative} for evaluating the
  degree of inflaton-mediated energy transfer between two radiation
  sectors: a less-equilibrated sector has a greater fraction of its
  particles concentrated in the region of phase space where resonant
  (and Bose-enhanced) inflaton-mediated scattering can occur. 
 
\item {\it The states in the radiation bath that dominate the energy
    dissipation from the inflaton are well-described as single
    (quasi-)particles.}  This is another manifestation of the implicit
  assumption that collective effects in the radiation bath can be
  largely neglected, and ensures that the description of reheating in
  terms of Boltzmann equations is valid. 

\end{itemize}

In contrast to many previous studies, however, {\it we use full
  Bose-Einstein or Fermi-Dirac distributions $f(p)$ to describe
  particles in thermal equilibrium.}  This is critical, as the
thermally averaged reaction rates for particles with different
statistics can have differences of multiple orders of magnitude.  We
give many new results on relativistic thermal averages in the
Appendices.

These simplifying assumptions allow us to expose the essential physics
without being overwhelmed by model-dependent details. While departures
from these assumptions will change the details of our conclusions, we
expect the general picture established in the following sections to be
preserved. 

\subsection{Thermal history of two-sector reheating }\label{sec:2secreheat}

We now develop the Boltzmann equations for two-sector reheating.  Our
primary aim is to investigate the possibility of thermalization
between the SM and a hidden sector in the limit where their only
non-gravitational interaction is the inevitable scattering via
inflaton exchange that follows from the assumption that both sectors
are reheated through couplings to the same field.  We take the
inflaton to have partial decay widths into two otherwise decoupled
sectors.  In general the two sectors can have different temperatures,
which we label as $T_a$ and $T_b$; where there is a distinction, we
will take $a$ to represent the SM and $b$ to represent the HS.

Two-sector reheating can be described by the coupled Boltzmann
equations
\begin {eqnarray}
\label{eq:master1}
\frac{d \rho_\phi }{dt}+3H\rho_\phi &=&-\Gamma_\phi\rho_\phi \\
\frac{d \rho_{SM}}{dt} +4H\rho_{SM} & = & \Gamma_{a}\rho_\phi  +\mathcal{C}
\\
\label {eq:master2}
\frac{d \rho_{HS}}{dt} +4H\rho_{HS} & = & \Gamma_{b}\rho_\phi  -\mathcal{C}  ,
\end{eqnarray}
where $\Gamma_{a} +\Gamma_{b} =\Gamma_{\phi}$, and the collision term
$\mathcal{C}$ describes the energy transfer between sectors that
results from inflaton-mediated $2\leftrightarrow 2$ scatterings.  We
consider the simple and illustrative case where the inflaton couples
to a single species of particle in each sector.  In this case, the
leading-order scattering matrix elements are completely determined by
the zero-temperature partial widths $\Gamma_{a,b}^0$ and $M_\phi$, and
have contributions in both $s$ and $t$ channels, 
\begin{eqnarray} 
\label{eq:coll1}
\mathcal {C}_s&=&-\int d\Pi_4 \left[   | \bar{\mathcal{M}} (12\to 34)| ^ 2 f_1f_2 (1\pm f_3)
  (1\pm f_4) ( E_1+E_2) -       \right. \\
\nonumber
&& \phantom {spacerrrrr} \left.  | \bar{\mathcal{M}} (34\to 12)| ^ 2 f_3f_4 (1\pm f_1)
  (1\pm f_2) ( E_3+E_4)\right] \\
\label {eq:coll2}
\mathcal {C}_t&=&-\int  d\Pi_4 \left[   | \bar{\mathcal{M}} (13\to 24)| ^ 2 f_1f_3 (1\pm f_2)
  (1\pm f_4) ( E_1-E_2) -       \right. \\
\nonumber
&& \phantom {spacerrrrr} \left.  | \bar{\mathcal{M}} (24\to 13)| ^ 2 f_2f_4 (1\pm f_1)
  (1\pm f_3) ( E_1-E_2)\right] ,
\end{eqnarray} 
where the two SM particles are labeled $1, 2$, the two HS
particles are labeled $3,4$, and $n$-particle phase space is defined
to include the internal degrees of freedom,
\begin{equation} 
d\Pi_n \equiv \prod_{i=1}^n \frac{g_i\, d^3\pp_i}{(2\pi)^3 2E_i} (2\pi)^4 \delta^4(\Sigma\,
p_i).
\end{equation} 

The inflaton decay rate $\Gamma_\phi$ also depends on the phase space
distributions $f (p)$ within each radiation sector: e.g.,
\begin{eqnarray}
\Gamma_a &=& \frac{1}{2M_\phi}\int  d\Pi_2 |\bar {\mathcal{M}}|^2 \left[(1\pm
  f_1)( 1\pm f_2) - f_1f_2 \right] \\
\label{eq:statgamma}
&=& \Gamma_a^0 \left( 1\pm 2 f(M_\phi/2) \right),
\end{eqnarray} 
where $\Gamma_a^0$ is the zero-temperature partial width, and in the
second line we have taken particles 1 and 2 to be described by a
common distribution $f (E)$.  Eqs.~\eqref{eq:master1}--\eqref{eq:master2}
are thus a full description of the system only when both sectors have
attained internal thermal equilibrium, and the $f_i$ can be taken to
be the equilibrium distributions in each sector; otherwise one must
also take into account the differential evolution of the $f_i$.

This brings us to another point: in eqs.\
\eqref{eq:master1}--\eqref{eq:master2}, we have also neglected the
production of finite-momentum inflaton quanta through their
interactions with the radiation baths.  In general these quanta will
be present, and should be included in a complete description of the
system.  We have omitted these interactions for simplicity, as even in
the cases where inflaton quanta are efficiently populated, they will
not affect our results, as we argue below.

We are interested in establishing the regime where the energy transfer
between sectors described by eqs.\ \eqref{eq:coll1}--\eqref{eq:coll2}
reaches equilibrium.  To do so, we consider the energy transfer rate
$\Gamma^E$ at $T_{RH}$, by which time we assume that the two sectors
have indeed attained internal thermal equilibrium, and compare it to
$H(\Trh)$.  If the energy transfer rate is in equilibrium for
temperatures $T<\Trh$, when inflaton decays cease to be important,
then any initial temperature asymmetry will have been erased, and no
subsequent asymmetry will be generated.

For this purpose, it suffices to consider $s$-channel scattering.  The
energy transfer in an $s$-channel process is maximal, the full
incoming energy that participates in the reaction: $\Delta E_s =
E_1+E_2$.  The energy transfer in a $t$-channel exchange is $\Delta
E_t = E_1-E_2$, and is comparable to the $s$-channel transfer only
when there is a substantial difference in temperatures between
sectors.  While $t$-channel scattering rates can be enhanced in the
forward region when $M_\phi \ll T$, the energy transfer in this region
is small.
The rates for $s$-channel scattering are also resonantly enhanced when
$T \sim M_{\phi}$, where $M_{\phi}$ is the mass of the inflaton.
Therefore, $s$-channel processes will give the dominant contribution
to the energy transfer, which simplifies our analysis. For the
remainder of our discussion we specialize to $s$-channel inflaton
exchange, as the inclusion of $t$-channel processes will not affect
our results.

%
\subsection{Determining $\Trh$}
\label{sec:reheatT}
%

As in the single sector case of~eq.\ \eqref{eqn:TRH1}, we estimate the
end of reheating as occurring when $\Gamma_\phi(T_a,T_b) = H(T_a,T_b)$, which
defines a reheat temperature in each sector.  Defining 
\be
T_b\equiv \xi T_a,
\ee 
this determines $T_{\mathrm{RH},a}$ in terms of the temperature
asymmetry $\xi$ through
\be
\label{eq:TRH_2sec}
\Gamma_\phi (\Trha, \xi) =  \sqrt{\frac{\pi^2 \tilde g (\xi)}{90}}    \frac{\Trha^2}{M_{\rm Pl}},
\ee
where we have defined
\begin{equation} 
\tilde g (\xi)\equiv g_{*,a}+ \xi^4 g_{*,b} .
\end{equation}

\begin{figure}[t]
\begin{center}\includegraphics[width=0.65\textwidth]{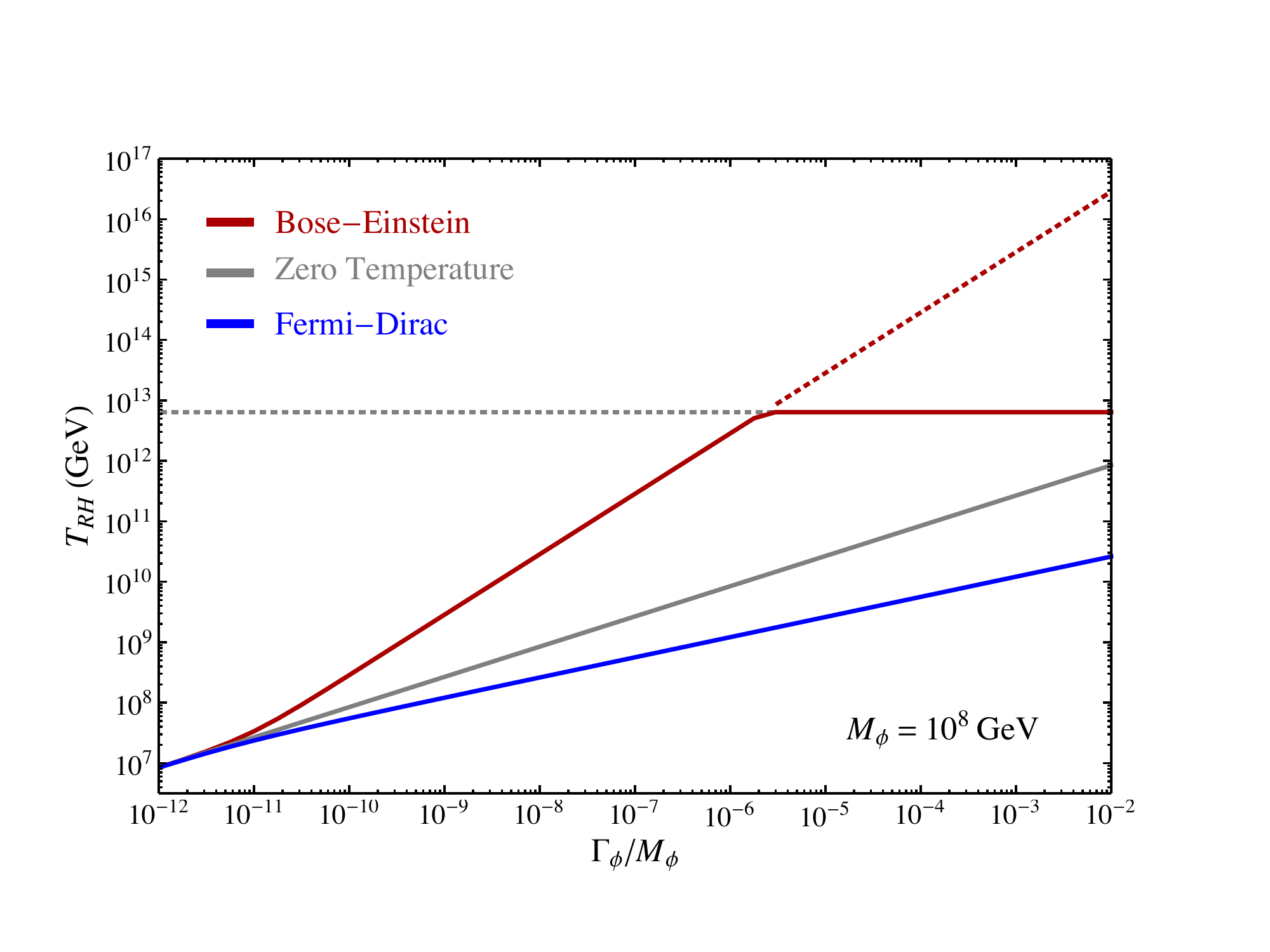}
  \caption{Estimates for (single-sector) $T_{RH}$ using Bose-Einstein
    (red) and Fermi-Dirac (blue) statistics, compared to the
    zero-temperature estimate (grey).  Here $M_\phi=10^8$ GeV.  The
    maximum achievable $\Trh$ is indicated by the grey dotted line;
    the estimate of eq.\ \eqref{eq:TRH_2sec} using Bose-Einstein
    statistics can exceed this for sufficiently strongly coupled
    inflatons, as indicated by the red dotted line. }
\label{fig:Trh}
\end{center}
\end{figure}

For the inflaton width, we use eq.\ \eqref{eq:statgamma} with equilibrium
phase space distributions in each sector, giving
\begin{equation}
\Gamma_{a,b}(T_{a,b}) = \Gamma_{a,b}^0  \left( 1\pm\frac{ 2}{e^{M_\phi/(2T_{a,b})} \mp 1}\right), 
\end{equation} 
where $\Gamma_{a,b}^0$ is the zero-temperature partial width, and the
upper (lower) sign holds for bosons (fermions).  In figure \ref{fig:Trh}
we show how quantum statistics affect the determination of $\Trh$ for
a single sector.  Pauli blocking and especially Bose enhancement have
a major impact for $\Trh \gtrsim M_\phi/2$, where $\Trh$ can differ by
orders of magnitude from the zero-temperature estimate.
The maximum achievable $\Trh$ corresponds to a near-instantaneous
transfer of energy from the oscillating scalar field to the radiation
bath, $T_{\mathrm{RH,max}}^2 \approx \sqrt{30/(\pi^2 \tilde g)}
M_\phi \Mp$.  The exponential enhancement of $\Gamma_\phi (T)$ in the
case of Bose statistics can result in extremely efficient inflaton
decays for temperatures sufficiently bigger than $M_\phi$, saturating
this upper bound.

To test whether inflaton scattering can thermalize the two sectors, we
assume that the two sectors have thermalized, i.e., we take $\xi = 1$,
and then check to see whether the resulting energy transfer rate can
exceed the Hubble rate at $T_{RH}$. If this criterion is met, then the
scenario is self-consistent, and thermalization can be achieved. To be
quantitative, we will need to assume a value for $g_{*,b}$. We take
for definiteness $g_{*, a}=g_{*, b}$, but our main results are not
sensitive to the details of this choice.  It is largely thanks to the
large SM value of $g_{*,a}$ that our results are insensitive to the
possible production of inflaton quanta: in the regions where
thermalization occurs and we expect inflaton quanta to be copiously
produced, the total amount of entropy carried by these inflaton quanta
will be an unimportant fraction of the entropy in the radiation bath.
Studies of low-scale reheating, where $g_\phi \ll g_{*,a} (\Trh)$ may
no longer hold, would need to more carefully revisit this point.

\subsection{Energy transfer rate}
\label{sec:Exfer}

We now turn to the collision term in eq.\ \eqref{eq:coll1}. Consider the
forward process in the $s$-channel collision term of
eq.\ \eqref{eq:coll1},
\begin{eqnarray}
\nonumber
n_{\rm eq}^2(T_a)\langle\sigma v \Delta E\rangle(T_a, T_b)&\equiv&
\int  \prod_{i=1}^4 d\Pi_i (2\pi)^4 \delta^4(\Sigma\,
p_i)  |\mathcal{M}(12\rightarrow34)|^2 (E_1+E_2) \times \\
\label{eqn:ExFer1}
&&
\phantom{move over} f_1(T_a)f_2(T_a)\,(1\pm f_3(T_b))\,(1\pm f_4(T_b)) ;
\end{eqnarray} 
the backward process can be written analogously as $n_{\rm
  eq}^2(T_b)\langle\sigma v \Delta E\rangle(T_b, T_a)$.  In thermal
equilibrium, the forward and backward processes are equal.

A convenient expression for the fractional energy transfer rate is
thus 
\be
\label{eq:gammaE}
\Gamma^E=\frac{(n^{\rm eq}(T_a))^{2}\langle\sigma v \Delta E\rangle(T_a, T_b)}{\bar{\rho}},
\ee
where $\bar \rho$ is the (average) energy density carried by the
incoming particle. When the particles that interact with the inflaton
in both sectors follow the same type of quantum statistics, we have
simply
\be
\bar \rho = g \frac{\pi^2}{ 30} T^4 \times \left\{ \begin {array} {cl}
    1& \mathrm{Bose},\\ 
    \frac{7}{8}  &\mathrm {Fermi}
    .\end{array}\right.  
\ee
When sectors $a$ and $b$
follow different quantum statistics, we define
\be
\bar \rho=\frac{\bar{\rho}^{BE}+\bar{\rho}^{FD}}{2}
\ee
in order to maintain manifest the invariance of the energy transfer rate
under exchanging sectors $a$ and $b$ when both sectors are in
equilibrium.

The evaluation of the integrals involved in the thermal average in
eq.\ \eqref{eqn:ExFer1} with Fermi-Dirac or Bose-Einstein distributions
is non-trivial, yet can be crucial for capturing the physical behavior
of relativistic scatterings. The majority of existing studies on
thermally averaged scattering cross-sections are in the context of
non-relativistic WIMP dark matter near its thermal freeze-out. In this
non-relativistic context, the phase space distribution is well
approximated by the Maxwell-Boltzmann distribution and phase space
blocking or enhancement factors can be neglected, both of which
greatly simplify the integration.    In the present case, we have no
reason to expect either of these approximations to hold.   
Here we adopt and extend the integration strategies originally
developed in \cite{YuehBuchler, Hannestad:1995rs} to study neutrino
decoupling in the early universe.  We conduct full computations with
both Fermi-Dirac and Bose-Einstein distributions for various
scattering cross-sections of interest.  The $s$-channel resonance
enhancement is found to be dramatic as long as there is sizeable phase
space where the center-of-mass energy of the particle collision is
around $M_\phi$ (i.e. $T$ is not too much lower than $M_\phi$). This
has a significant impact on thermalization between the two sectors. To
the best of our knowledge, many of the results presented here based on
full treatment of relativistic scatterings are new to the literature,
and highlight the critical importance of treating quantum statistics
accurately at high temperature, particularly in the case of scalars.
In order to not distract from the physics, we present the somewhat
gory details of our calculations in appendix~\ref{App:Exfer}.

\section{Inflaton-mediated thermalization in simple models of two-sector reheating}
\label{sec:examplesReh}

We are now ready to use the set-up of section \ref{sec:2secreheat} to conduct
quantitative studies of several simple representative models of two-sector
reheating that cover various assignments of quantum numbers to the
inflaton and its decay products.  For each example model we consider,
we derive an approximate analytic expression for the energy transfer
rate $\Gamma^E$, then check to see in what region of parameter space
the self-consistent thermalization condition
\begin{equation} 
\Gamma^E (T) > H(T),
\end{equation} 
can be satisfied for $T\leq \Trh$.  In most cases, $\Gamma^E(T)$
decreases faster with $T$ than does $H(T)$. This means that if
thermalization does not happen by $\Trh$, it will not occur at lower
temperatures.  However, there are two exceptions, as we will discuss
in more detail in the following subsections. First, the resonant
structure of the energy transfer rate becomes important for $T\sim
2M_\phi/5$, which can allow thermalization even if
$\Gamma^E(\Trh)<H(\Trh)$.  Second, in the case where the inflaton and
its decay products in both sectors are all scalars,
$\Gamma^E(T)\propto T$ for $T\ll M_\phi$, making thermalization
possible at late times.
  
We consider fixed example values of $M_\phi$, and for simplicity, we
assume all external masses $m \ll M_\phi, \,\Trh$ to be equal.  In
practice, we take a small but non-zero value of 10 GeV for these
masses to avoid numerical artifacts, unless otherwise indicated.  In
our scan of parameter space we parametrize the models using the
variables
\begin{align}
w\equiv&\frac{\Gamma^0_{a}+\Gamma^0_{b}}{M_\phi},\quad
k^2\equiv\frac{\Gamma^0_{b}}{\Gamma^0_{a}},
\end{align}
where we generally assume $k\leq 1$.  The parameter $w$ represents the
(zero-temperature) inflaton decay width relative to its mass and is
related to $\Trh$ via eq.\ \eqref{eq:TRH_2sec}, while the parameter
$k$ characterizes the ratio of zero-temperature widths to SM and HS
states. To better illustrate the physics in variables more closely
related to observables, we will present our results in the plane of
$(k,\Trh)$, where $\Trh$ has a one-to-one correspondence with $w$ for
a given $M_\phi$.

To be as model-independent as possible, we consider a general range of
possible values for $w$.  We restrict $w\lesssim 0.01$ for
perturbativity, which can limit the achievable $\Trh$ in specific
models. Among the models we consider is the case where the inflaton
has axion-like couplings to gauge bosons and fermions via
dimension-five operators, in which case the validity of the effective
field theory (EFT) imposes limits on the achievable values of $w$ for
a given $M_\phi$.

We derive analytic approximate formulae for the thermally averaged
energy transfer rate and apply them in our numerical scan. The details
of these approximations are presented in
appendix~\ref{App:specifics}. Our analytic formulae agree well with
the results from exact numerical evaluation from $T\gg M_\phi$ down to
$T\sim m$, where $m$ is the external particle mass.  For temperatures
below $T\sim m$, we expect that the Boltzmann suppression of the
external states will render inflaton-mediated scattering unimportant
on cosmological timescales.

\subsection{Scalar inflaton  coupling to scalar pairs in both sectors}
\label{sec:scalartherm}

We first consider the model where the inflaton is a CP-even scalar
which couples to scalars in both sectors via dimensionful trilinear
couplings. The relevant Lagrangian in this case is
\be
\label{eq:scalarLag1}
\mathcal{L}\supset - \frac{1}{2}\mu_a\phi S_a^2 -  \frac{1}{2}\mu_b\phi S_b^2,
\ee
where we have not explicitly written the mass terms for the fields,
taking $m_{S_a}=m_{S_b}\equiv m_S\ll M_\phi$.  In this minimal model
we take the $S_i$ to be real singlets; extensions to complex scalars
(such as the SM Higgs) are trivial.  The zero-temperature partial
widths are thus
\beq
\label{eq:scalarLag2}
\Gamma^0_{a,b}=\frac{1}{32\pi}\frac{\mu_{a,b}^2}{M_\phi}\sqrt{1-\frac{4m_{S}^2}{M_\phi^2}},
\eeq
yielding $k^2= \mu_b^2/\mu_a^2$. We can express $\mu_a$
in terms of $k$ and $w$ as 
\begin{equation} 
\mu_a^2=\frac{w}{1+k^2}\frac{32\pi M_\phi^2}{\sqrt{1-\frac{4m_S^2}{M_\phi^2}}}.
\end{equation} 
The amplitude for $S_aS_a\leftrightarrow S_bS_b$ scattering is given by
\be
\label{eq:scalarLag3}
|\widebar{\mathcal{M}}(s)|^2 =\frac{\mu^2_a \mu^2_b}{(s-M_\phi^2)^2 + M_\phi^2(\Gamma_\phi^0)^2} .
\ee

\begin{figure}[t]
\begin{center}
\includegraphics[width=\textwidth]{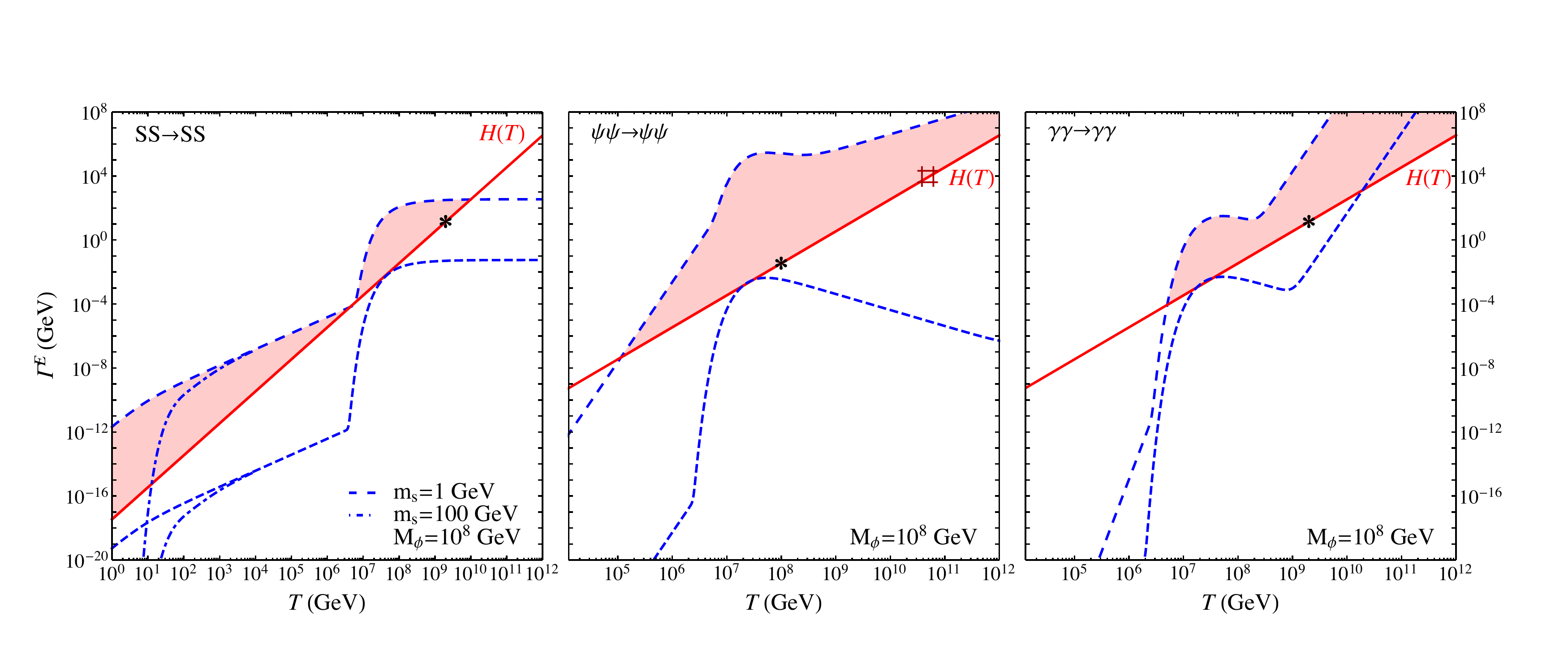}
\caption{Comparison of the energy transfer rate (blue, dashed) to the
  Hubble rate (red, solid) for the scalar-scalar model (left panel),
  the Dirac fermion-Dirac fermion model (middle panel) and the
  gauge-gauge boson model (right panel). In each case the energy
  transfer rate is shown for a value of $k = 0.5$ and for $M_{\phi} =
  10^8$ GeV. The lower curve in each case is the lowest value of the
  width for which thermalization occurs (for $k = 0.5$), while the
  upper curve is the upper allowed value. The black star indicates the
  reheat temperature and corresponding Hubble rate for the lower
  width, while the red \# (visible only in the fermion case) indicates
  the maximum value of the reheat temperature.}
\label{fig:eg1}
\end{center}
\end{figure}

\begin{figure}[t]
\begin{center}
\hskip - .5cm \includegraphics[width = \textwidth]{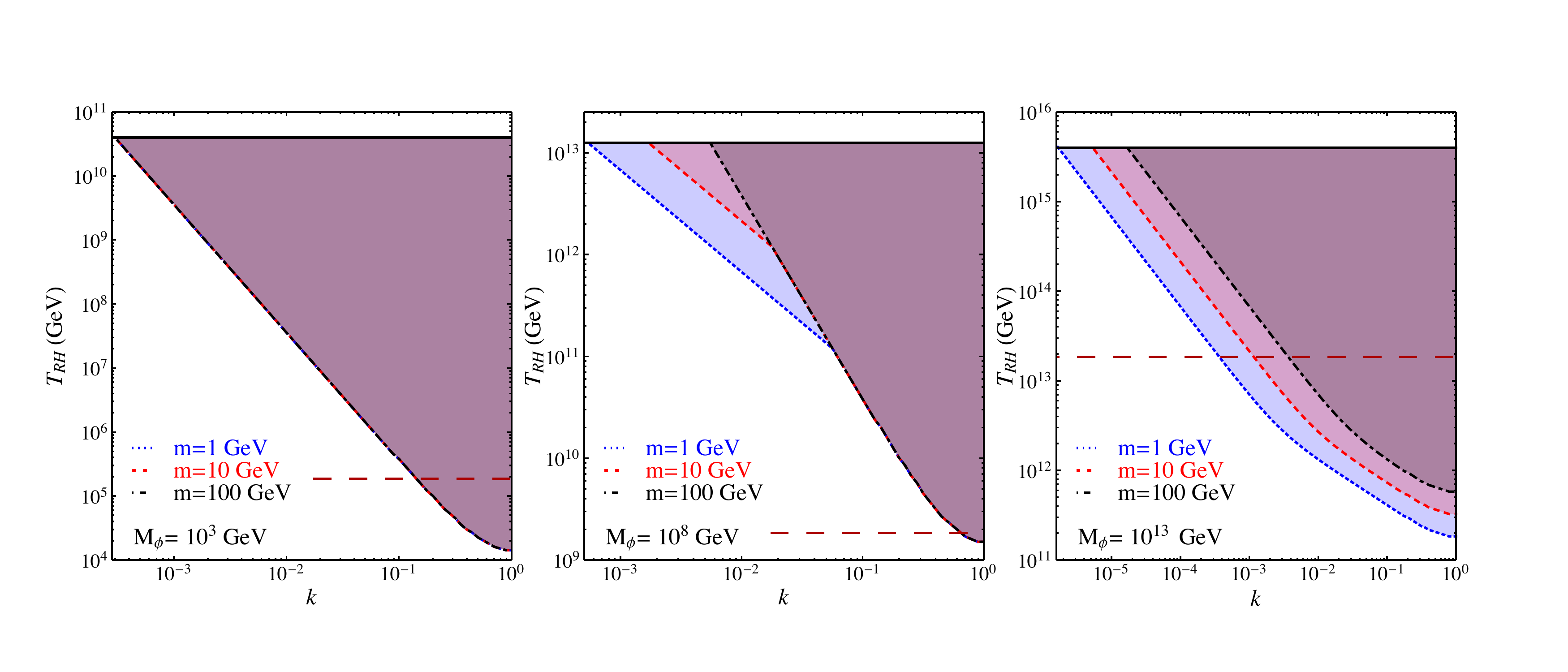}
\caption{Shaded regions indicate the parameter space where
  inflaton-mediated scattering can thermalize two otherwise decoupled
  sectors, in the case where a scalar inflaton couples to scalars in
  both sectors, for $M_\phi = 10^3$ GeV (left), $10^8$ GeV (center),
  and $10^{13}$ GeV (right). Results are shown for three different
  choices of external scalar mass.  Above the horizontal dashed red
  line, preheating through broad resonance is always relevant.
}
\label{fig:eg2}
\end{center}
\end{figure}

The derivation of the energy transfer rate for this case can be found
in appendix~\ref{sec:scaltrilin}.  More than any other case, the
potentially resonant scattering of scalars to scalars demonstrates the
importance of retaining full dependence on the quantum statistics in
evaluating the thermal average.  The energy transfer rate is plotted
as a function of temperature in the left panel of figure  \ref{fig:eg1}
(see also figure \ref{fig:scalartrilinear}) for two different values of
the width. These values correspond to the boundaries of the middle
panel of figure  \ref{fig:eg2} at $k = 0.5$.  Strikingly, the rate
asymptotes to a constant at high temperature, thanks to the
non-vanishing overlap of the zero-momentum singularity in the
Bose-Einstein distribution with the pole in the scattering amplitude.

Scattering in the scalar case has the unusual feature that, as it is
controlled by super-renormalizable couplings $\mu_{a,b}$, it falls off
more slowly than does the Hubble rate as the temperature
decreases. Below the scale of the inflaton mass, we can estimate
\begin{equation} 
\label{eq:gammaEapproxscalar}
\Gamma^E (T<M_\phi) \sim \frac{\mu_a^2\mu_b^2}{32\pi M_\phi^4} T,
\end{equation} 
while the Hubble rate decreases as $H\propto T^2$ during the
radiation-dominated era. We compare the evolution of the two rates as
a function of temperature for one choice of parameters in the left
panel of figure \ref{fig:eg1}.  Thus in this particular case,
thermalization can be controlled by mass scales in the infrared: in
our simple models, the lowest temperature where thermalization is
possible is set by the mass scale of the external states, $T\sim m_S$.
In this model, when we check for thermalization, we check to see
whether $\Gamma^E(T)>H(T)$ for {\em any} $T$ in the range $(m_S,
\Trh)$.  We continue to take $\xi=1$ in determining $\Trh$, even
though thermalization may not occur until much lower
temperatures. This is conservative, as keeping $\xi<1$ earlier in
the radiation-dominated era would decrease the Hubble rate relative to
$\Gamma^E$, thereby making thermalization easier.

The region of parameter space where thermalization can be achieved in
this model is shown in figure \ref{fig:eg2} for three reference choices
of the inflaton mass.  The maximum $\Trh$ shown in these plots is the
maximum allowed by energy conservation, i.e., is the value realized by
instantaneous reheating; the tree-level couplings are still
perturbative.

As is evident, thermalization preferentially occurs for high values of
$\Trh$, despite the IR sensitivity in this model.  This can be
understood simply from dimensional analysis. Using the estimate of
eq.\ \eqref{eq:gammaEapproxscalar} and $T_{RH}\sim \sqrt{\Gamma_\phi
  \Mp}$ and dropping numerical constants, the criterion for
thermalization becomes
\begin{equation} 
\frac{T}{\Trh} < k^2 \frac{\Trh}{\Mp} \,\left(\frac{\Trh}{M_\phi}\right) ^ 2 .
\end{equation} 
Thus we can make two important observations: first, that late-time
thermalization will be more important in theories that yield high
$\Trh$, and second, that thermalization is easiest when $ \Trh > 
M_\phi$.  This estimate neglects the possible Bose enhancement in
determining $\Trh$ (and in scattering at $\Trh$); this enhancement
increases $\Trh$ for fixed $M_\phi$, $\Gamma_\phi^0$, further reinforcing
these two observations.

In figure \ref{fig:eg2}, we observe that for $M_\phi = 10^3$ GeV,
thermalization occurs near the resonance, and consequently the
thermalizing parameter space does not depend on the value of the
external $m_S$.  The energy transfer rate drops sharply below the
resonance, and there are not enough decades of temperature in the
remaining range $m_S < T < M_\phi$ to allow the energy transfer rate
to come back into equilibrium.  Meanwhile for $M_\phi = 10^ {13}$ GeV,
$m_S \ll \Trh$, and if thermalization occurs, it does so in the IR.
The value $M_\phi = 10^8$ GeV realizes an intermediate case, with lower
values of $\Trh$ yielding thermalization at $T\sim M_\phi$, while
higher values of $\Trh$ enable late-time thermalization.

The regions where thermalization can occur are characterized by
relatively strong couplings of the inflaton to matter. This is
precisely the region of parameter space where we expect
non-perturbative particle production --- i.e., preheating --- to
potentially be important.  For sufficiently large couplings of the
radiation bath to the inflaton, the time-dependence of the oscillating
inflaton background will lead to particle production.  We would
typically expect non-perturbative particle production to become
important when the background-induced time variation of the frequency
for the radiation modes is no longer small relative to the frequency:
\begin{equation}
\label{eq:adiabatic}
\dot \omega \gtrsim \omega^2 ,
\end{equation} 
which, for our trilinear model, is parametrically the condition that
$\mu\Phi \gtrsim M_\phi^2$.
Note that this condition depends not only on the inflaton mass and
couplings to radiation, but also the amplitude of the collective
oscillations.  Thus specifying $\Gamma_\phi$ and $M_\phi$ is not a
priori enough information to determine the importance of preheating.
However, the Hubble parameter during reheating is bounded: on one
hand, reheating occurs for $H\gtrsim \Gamma_\phi$, and on the
other hand, we must have $H\lesssim M_\phi$ if $\phi$ is not to
inflate (we assume, as always, that the scalar field dominates the
energy density of the universe and that the region of the potential
explored by $\phi$ is quadratic).  This translates parametrically into
a finite possible range for $\Phi$,
\begin{equation} 
\label {eq:phirange}
\frac{\Gamma_\phi}{M_\phi}\lesssim \frac{\Phi}{\Mp} \lesssim 1.
\end{equation} 
As we review in appendix~\ref{sec:preheating_scalars}, the regions in
parameter space where preheating is relevant for scalar trilinear
couplings are largely determined by the value of the parameter
$q\equiv 2\mu\Phi/M_\phi^2$.  The violation of adiabaticity of eq.\ 
\eqref{eq:adiabatic} corresponds to the condition that $q>1$, which
defines the ``broad resonance'' regime.  Here preheating can yield a
very efficient transfer of energy from the inflaton condensate to the
radiation bath, resulting in the universe spending little to no time
in the perturbative oscillatory phase.  Thus, when the broad resonance
regime pertains, we obtain a higher formal value for $\Trh$ given
fixed inflaton mass and couplings than the perturbative estimate would
indicate.  A higher value of $\Trh$ in turn translates into more
efficient thermalization between the two sectors. Of course, we expect
that sectors preheated via a broad resonance may not have had time to
attain internal thermal equilibrium by the end of reheating.  This
will also tend to enhance inflaton-mediated scattering, as the regions
of phase space with momentum $p\sim M_\phi/2$ will be preferentially
occupied relative to the equilibrium distribution.

Given $\Gamma_\phi$ and $M_\phi$, for any condition on $q$, we can
thus identify three regions: (i) where the condition cannot be met for
any $\Phi$ in the allowed range of eq.\ \eqref{eq:phirange}; (ii) where
the condition can be met for some but not all allowed $\Phi$; and
(iii) where the condition is met for all allowed values of $\Phi$.  In
figure \ref{fig:eg2} we have indicated how these regions for the broad
resonance condition $q>1$ intersect the parameter space that realizes
thermalization.  Above the dashed red line, broad resonance is
operative for all $\Phi$, while below this line, broad resonance is
realized for some $\Phi$.

\subsection{Scalar inflaton coupling to Dirac fermion pairs in both
  sectors}
\label{sec:fermitherm}

We next consider the case where a scalar inflaton has renormalizable
couplings to a pair of fermions in each sector. For simplicity, we
take these fermions to be Dirac, but results for Majorana fermions
(such as, e.g., right-handed neutrinos in the SM) are very similar.
The relevant Lagrangian in this case is
\be
\label{eq:fermiLag1}
\mathcal{L}\supset - y_a\phi\bar{\psi}_a\psi_a - y_b\phi\bar{\psi}_b\psi_b,
\ee
yielding the zero-temperature partial width 
\be
\label{eq:fermiLag2}
\Gamma^0_{a,b}=\frac{y_{a,b}^2 M_\phi}{8\pi}\left(1-\frac{4m_{\psi}^2}{M_\phi^2} \right)^{3/2} .
\ee
The summed, averaged amplitude for
$\bar{\psi}_a\psi_a\leftrightarrow\bar{\psi}_b\psi_b$ scattering is
\be
\label{eq:fermiLag3}
\left|\widebar{\mathcal{M}}(s)\right|^2 = y_a^2y_b^{2}\left(1-\frac{4m_\psi^2}{s}\right)^2\frac{s^2}{(s-M_\phi^2)^2 + M_{\phi} ^2 (\Gamma_{\phi}^0)^2}.
\ee
The full expression for the energy transfer rate for this case is
derived in appendix~\ref{sec:fermifermi}, and plotted in
figure \ref{fig:fermion-yukawa}.  In this case, Maxwell-Boltzmann
distributions are not an unreasonable guide to the behavior of the
thermal average, giving results that vary by only factors of order
unity from the exact result. We use the approximate analytic
expression to the Fermi-Dirac result of eq.\ 
\eqref{eqn:fermionyukawaFDapp} in our numerical scans. In the middle
panel of figure \ref{fig:eg1} we plot the temperature dependence of the
energy transfer rate for this model compared to the Hubble rate.


\begin{figure}[t]
\begin{center}
\includegraphics[width = \textwidth]{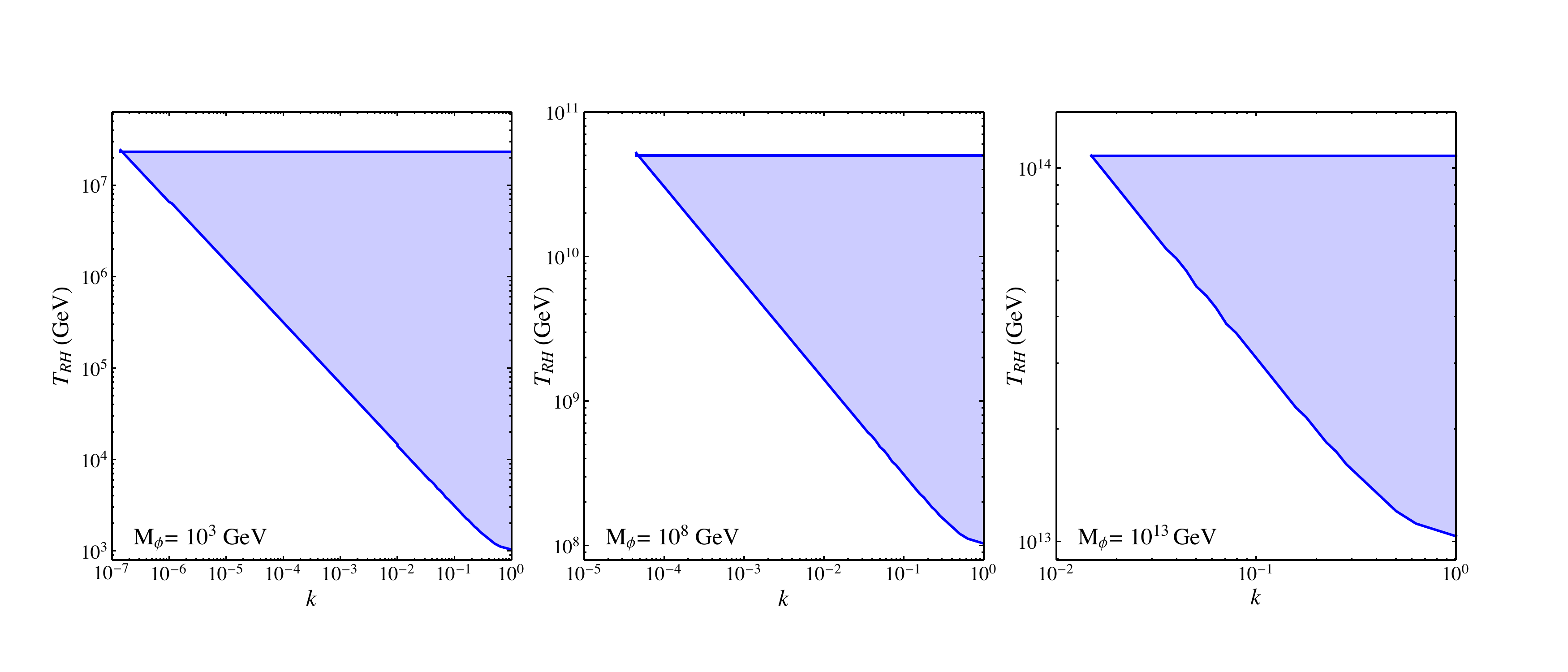}
\caption{Shaded regions indicate the parameter space where
  inflaton-mediated scattering can thermalize two otherwise decoupled
  sectors, in the case where a scalar inflaton couples to fermions in
  both sectors, for $M_\phi = 10^3$ GeV (left), $10^8$ GeV (center),
  and $10^{13}$ GeV (right). The upper boundary on these plots is
  determined by the maximum reheat temperature that can be obtained
  with $w < 0.01$.  }
\label{fig:eg6}
\end{center}
\end{figure}

In figure \ref{fig:eg6} we show the parameter space where
thermalization can be achieved in this model. We cut the parameter
space off when the width exceeds $w > 0.01$. Note that the maximum
$\Trh$ obtained in these models is smaller than in the previous case
with final state scalars: this highlights the impact of Pauli blocking
at $\Trh >M_\phi$.  It is immediately obvious from the figure that in
this case thermalization requires $\Trh > M_\phi$.  Here, we can
estimate (dropping numerical factors) that, neglecting resonant
effects, thermalization at $\Trh$ requires
\begin{equation} 
1\sim k^2 \frac{\Trh}{\Mp}\, \left(\frac{\Trh}{M_\phi}\right) ^6
\end{equation} 
which cannot be satisfied for $\Trh < M_\phi$.  For $\Trh > M_\phi$,
the availability of the resonance helps boost the energy transfer rate
even above the estimate from dimensional analysis, and can enable the
two sectors to thermalize.

For fermions, Pauli blocking prevents non-perturbative particle
production from being an efficient dissipation mechanism for the
inflaton condensate, regardless of the strength of the coupling $y$ or
the amplitude of oscillation $\Phi$, as we review in
appendix~\ref{sec:preheating_fermions}.

\subsection{Pseudo-scalar inflaton coupling to gauge boson pairs in
  both sectors}
\label{sec:gaugetherm}

The third case we consider is a pseudo-scalar inflaton with axion-like
couplings to gauge bosons in both sectors.  We consider Abelian gauge
fields for simplicity; the generalization to non-abelian cases is
straightforward. The relevant Lagrangian is
\be
\label{eq:gaugeLag1}
\mathcal{L}\supset -
\frac{1}{4\Lambda_a}\phi \, 
F_a^{\mu\nu}\widetilde F_{a\mu\nu} - \frac{1}{4\Lambda_b}\phi\, 
F_b^{\mu\nu}\widetilde F_{b \mu\nu} 
\ee
The zero-temperature decay width to each sector is then 
\be
\Gamma_{a,b}^0=\frac{M_\phi^3}{256\pi \Lambda_{a,b}^2} \left( 1-\frac{4m_\gamma^2}{M_\phi^2}\right)^{3/2}
\label{eq:width_axion}
\ee
where we have included possible finite photon masses from e.g. thermal
effects.  The summed, averaged amplitude for
$\gamma_a\gamma_a\leftrightarrow \gamma_b\gamma_b$ scattering is then
\be
\label{eq:gaugeLag3}
|\widebar{\mathcal{M}}(s)|^2 =
\frac{1}{128\Lambda^2_a\Lambda_b^2}\left( 1-\frac{4m_\gamma^2}{M_\phi^2}\right)^2\frac{s^4}{(s-M_\phi^2)^2
  + M_{\phi} ^2 (\Gamma_{\phi}^0)^2} .
\ee
In order for this EFT description of the axion interactions to make
sense, we require $M_\phi<\Lambda_i$.  We will also require
$\Trh<\Lambda_i$ in our numerical scans over parameter space, in order
to maintain the validity of the EFT in describing scattering at high
temperatures.  Both conditions place nontrivial restrictions on the
zero-temperature mass-to-width ratio $w$.

We discuss the derivation of the resulting energy transfer rate in
appendix~\ref{sec:gaugegauge}.  In this case, the full use of
Bose-Einstein statistics does not lead to dramatic enhancements in the
energy transfer rate at high temperatures. This occurs because the
non-renormalizability of the inflaton couplings suppresses
contributions to the scattering amplitude from low momenta, with the
consequence that Maxwell-Boltzmann is a fairly good approximation to
the full result, as shown in figure \ref{fig:gauge-axion} (see also the
right panel of figure  \ref{fig:eg1}).  The analytical approximation we
use in our scans is given in eq.\ \eqref{eqn:gaugembenergyxferBE}.


\begin{figure}[t]
\begin{center}
\includegraphics[width = \textwidth]{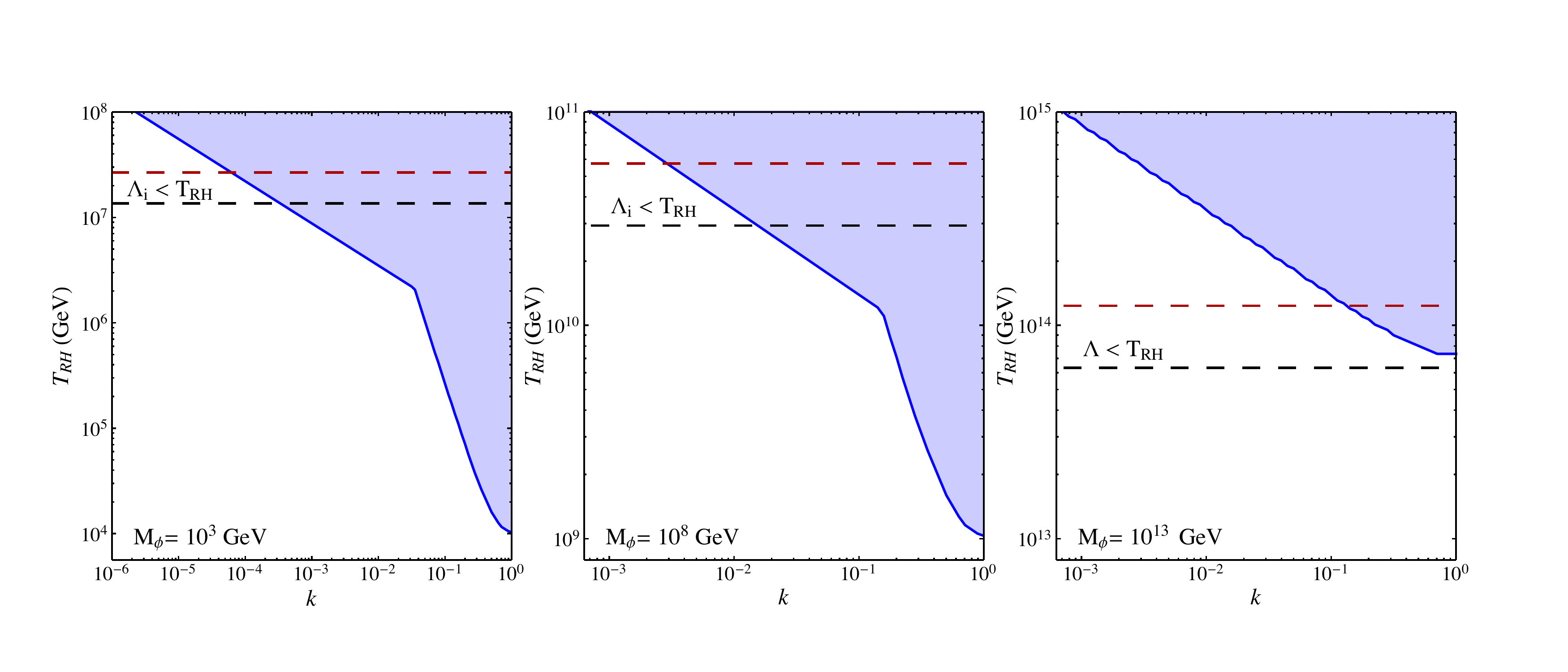}
\caption{Shaded regions indicate the parameter space where
  inflaton-mediated scattering can thermalize two otherwise decoupled
  sectors, in the case where a pseudo-scalar inflaton couples to gauge
  bosons in both sectors, for $M_\phi = 10^3$ GeV (left), $10^8$ GeV
  (center), and $10^{13}$ GeV (right).  Above the black dashed line,
  the EFT description of the inflaton couplings breaks down, while
  above the red dashed line preheating is always important.}
\label{fig:eg4}
\end{center}
\end{figure}

In figure \ref{fig:eg4} we show the region of parameter space where
thermalization can occur in this model.  We cut the parameter space
off somewhat above the EFT bound to show context; for instance, in the
case of $M_\phi = 10^{13}$ GeV, thermalization is not possible for
$\Trh < \Lambda_i$.  Thermalization in models with lower values of
$\Trh$ is dominated by the resonant enhancement at $T\lesssim M_\phi$,
while in models with higher values of $\Trh$, thermalization occurs
for $T\sim \Trh$. This is evident from the right panel of figure 
\ref{fig:eg1}. In the lower curve in this figure (corresponding to the
lowest value of the width that equilibrates at $k = 0.5$),
thermalization occurs at $T < \Trh$ as scattering becomes resonant
following reheating. As the width, and thus reheating temperature, is
increased, the energy transfer rate can exceed the Hubble rate at
reheating itself. However, thermalization cannot occur unless $\Trh
\gtrsim M_\phi$, as again can be supported by dimensional analysis.

In this model, non-perturbative particle production can again become
the dominant process governing the decay of the inflaton condensate
for sufficiently large couplings to matter $\Lambda$ and sufficiently
large field oscillations $\Phi$, as reviewed in
appendix~\ref{sec:preheating_axion}.  The red dashed line in
figure \ref{fig:eg4} shows the boundary between the regime where broad
resonance preheating is important for any physical $\Phi$, and the
regime where broad resonance preheating is important for some but not
all physical $\Phi$.  All thermalizing regions within the range of
validity of the EFT lie in the region where preheating can be, but is
not necessarily, realized.  As for scalars, we expect that when
preheating is important, it will tend to enhance thermalization,
thanks to both the higher values of $\Trh$ and the imperfect
equilibration of the radiation sectors.

\subsection{Hybrid cases}  
\label{sec:hybrid}

Finally, we consider cases where the particles that couple to the
inflaton in the hidden and visible sectors follow different quantum
statistics.  We consider both the case of a scalar inflaton coupling
to both fermions and scalars, and a pseudo-scalar inflaton coupling to
fermions and gauge bosons.

\subsubsection{Fermion-scalar scattering via scalar exchange }

The relevant Lagrangian in this case is
\be
\label{eq:hybridLagA1}
\mathcal{L}\supset -\frac{1}{2}\mu\phi S^2-y\phi\bar{\psi}\psi,
\ee  
where again for simplicity we take the scalar, $S$, to be real and the
fermion, $\psi$, to be Dirac.
Considering $S$ to be in sector $a$ and $\psi$ in sector $b$, this
translates into our scan parameters as
\begin{align}
\label{eq:hybridLagA2}
   \mu=&4\sqrt{\frac{2\pi w}{1+k^2}}M_\phi,
   \quad y=\frac{\mu k}{2M_\phi} ,
\end{align}
where $k$ can now take on any real value (here we have dropped the
small external masses).  The summed, averaged amplitude for $SS\to
\bar{\psi}\psi$ scattering is
\be
\label{eq:hybridLagA3}
|\widebar{\mathcal{M}}(s)|^2 = \frac{2 \mu^2y^2  s}{(s-M_{\phi}^2)^2 +
  M_{\phi}^2 (\Gamma_\phi^0)^2}\left(1-\frac{4m_\psi^2}{s}\right) .
\ee
The full expression for the energy transfer rate for this case is
derived in appendix~\ref{sec:fermion-scalar}, and plotted in
figure \ref{fig:fermionscalar}.  We use the approximate analytic
expression of eq.\ \ref{eqn:FSFDBEapprox} in our numerical scans.

 \begin{figure}[t]
\begin{center}
\includegraphics[width=\textwidth]{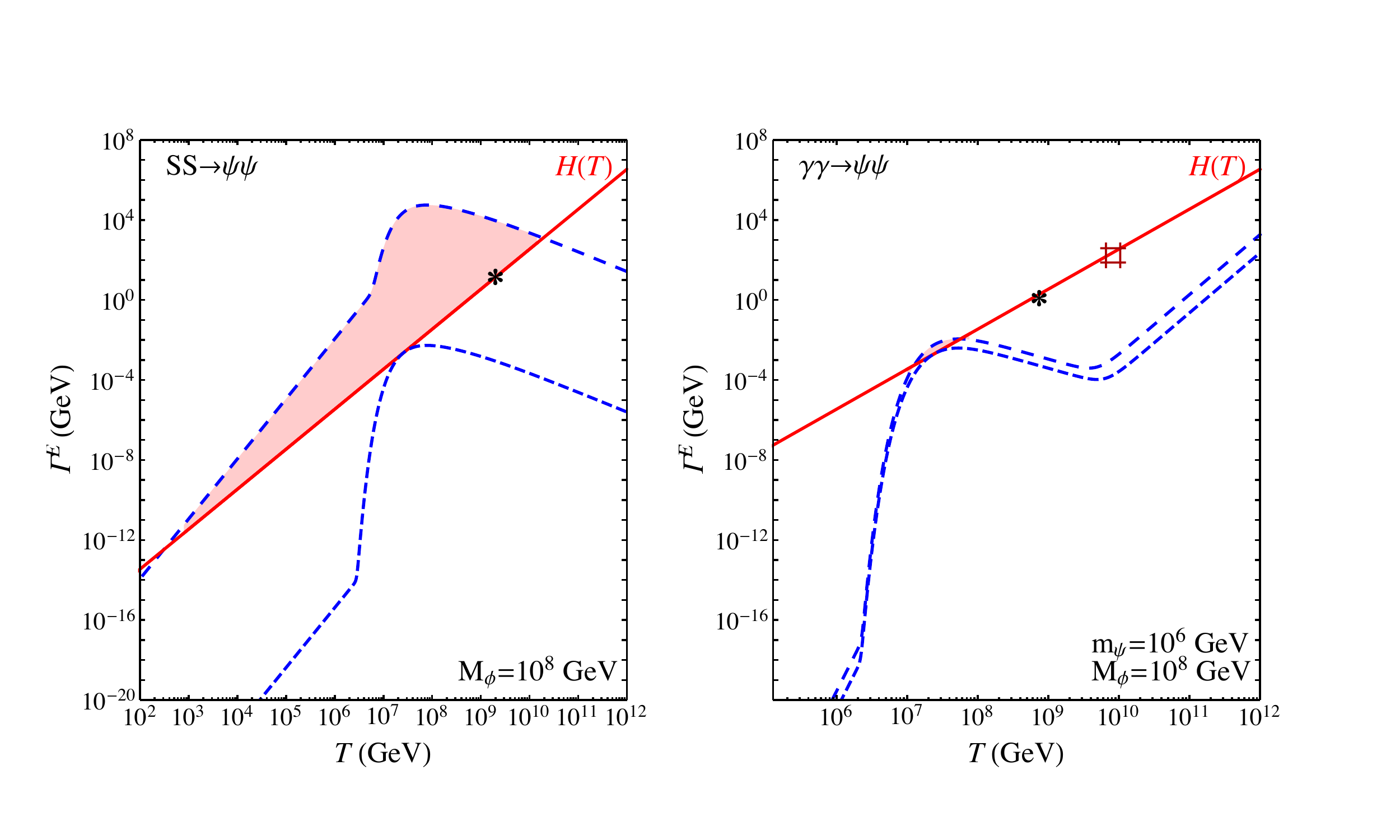}
\caption{Rate comparison for the hybrid cases of fermion-scalar
  scattering (left panel), and fermion-gauge scattering (right
  panel). In each case, plotted is the energy transfer rate at the
  lowest value of the width that thermalizes (lower curve), and the
  maximum allowed value of the width (blue, dashed). For the
  fermion-scalar case, we show the rate for $k = 2$, while for the
  fermion-gauge case we show the rate for $k = 0.5$. We also show the
  Hubble rate (red, solid). The black asterisk ${\bf *}$ indicates the
  reheat temperature corresponding to the lowest width that
  thermalizes, while the red $\#$ indicates the maximum value of the
  reheat temperature.}
\label{fig:eg9}
\end{center}
\end{figure}

  \begin{figure}[t]
\begin{center}
\includegraphics[width = \textwidth]{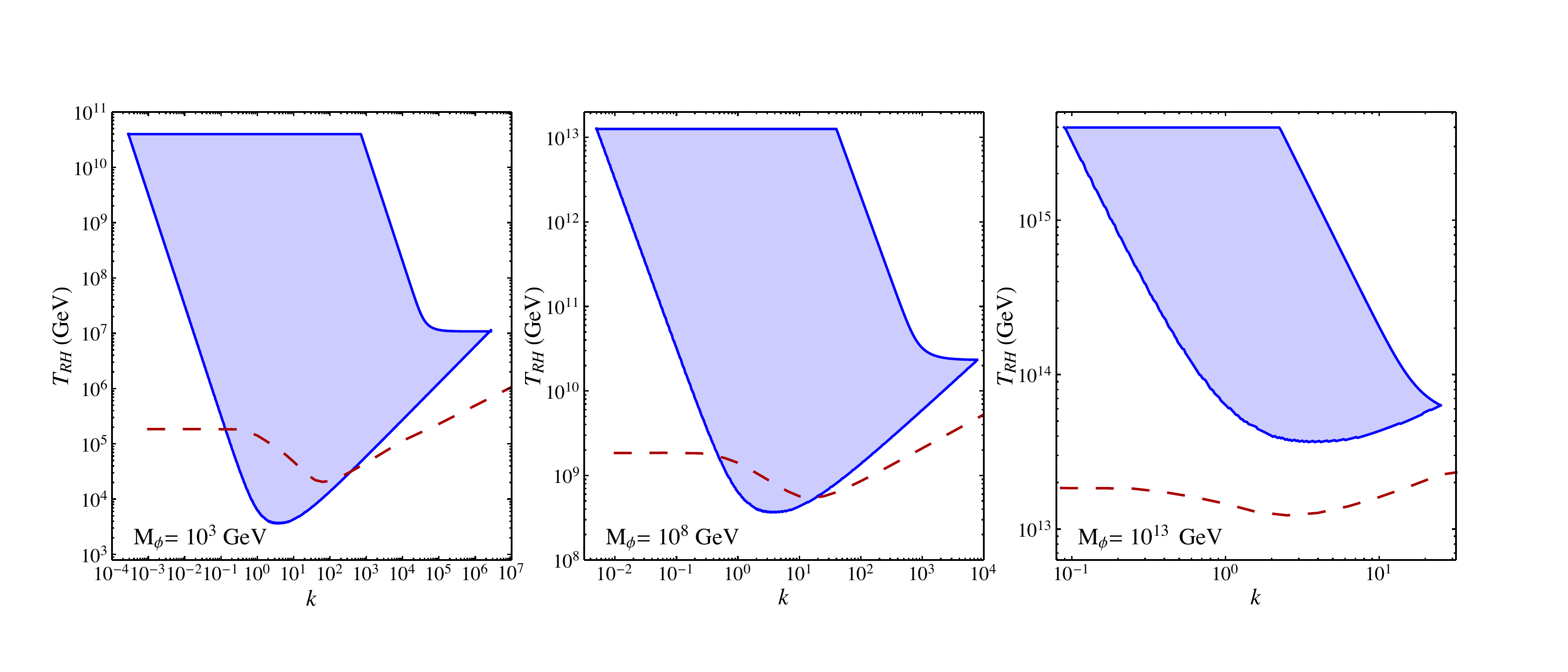}
\caption{Parameter region where thermalization can be achieved for the
  scalar-fermion model for $M_\phi = 10^3$ GeV (left), $10^8$ GeV
  (center), and $10^{13}$ GeV (right). The upper boundaries of these
  regions are set by the maximum attainable reheating temperature in
  each case. For fermion-dominated decays, the upper limit is set by
  perturbativity, which limits $w < 0.01$. For boson-dominated decays,
  conservation of energy in instantaneous reheating fixes the upper
  limit. The red dashed line indicates the lower boundary of the
  region where preheating is always important.}
\label{fig:eg10}
\end{center}
\end{figure}

As in our previous examples, dimensional analysis indicates that
thermalization of the two sectors through inflaton exchange requires
$\Trh > M_\phi$.  The region where thermalization occurs is shown in
figure \ref{fig:eg10}.
As indicated by the red dashed line, the parameter space that realizes
thermalization will also generically undergo broad resonance
preheating, with the exception of a limited parameter space at lighter
inflaton masses where preheating can but does not necessarily occur.

\subsubsection{Fermion-gauge boson scattering via pseudo-scalar exchange}

Here we consider a pseudo-scalar inflaton that couples to Dirac
fermions in one sector, and (Abelian) gauge bosons in another sector.
The relevant Lagrangian is then given by
\be
\label{eq:hybridLagB1}
\mathcal{L}\supset -
\frac{m_\psi}{\Lambda_b}\phi\bar{\psi}\gamma_5\psi- \frac{1}{4\Lambda_a}\phi\, F^{\mu\nu}\widetilde{F}_{\mu\nu}.
\ee
We have parameterized the fermion-pseudoscalar coupling in terms of a
chiral-symmetry-breaking mass $m_\psi$, as generally in realistic
models this interaction is effective dimension five; we thus typically
expect the effective Yukawa coupling $m_\psi/\Lambda_b \ll
1$.\footnote{For this reason, we have not considered the case where a
  pseudo-scalar inflaton mediates fermion-fermion scattering.  Should
  this scenario become of interest, in the relativistic regime under
  consideration the results are in any case identical to those for a
  scalar inflaton, given in section \ref{sec:fermitherm}.}  The
inflaton's zero-temperature partial width to fermions is then
\begin{align}
\label{eq:hybridLagB2}
\Gamma^0 (\phi\rightarrow\bar{\psi}\psi)=&\frac{1}{8\pi}\frac{m_\psi^2
M_\phi}{\Lambda_b^2}\sqrt{1-\frac{4m_\psi^2}{M_\phi^2}},
\end{align}
giving
\begin{align}
\Lambda_a=\frac{1}{16}\sqrt{\frac{1+k^2}{\pi w}}M_\phi, \quad 
\Lambda_b=m_\psi \sqrt{\frac{1+k^2}{8\pi w k^2}}
\end{align}
(we again here neglect subleading dependence on the external
masses). In principle, $k$ can take any value, but as we typically
expect gauge couplings to dominate reheating when they are available,
we will consider only $k<1$ in our scans. We fix $m_\psi = 10^{-2}
M_\phi$ and allow $\Lambda_b$ to vary, subject to $\Lambda_b > \Trh,
M_\phi$.  This necessitates using an equal external mass for the gauge
bosons, as our analytic approximation to the energy transfer rate
holds in the limit that all external masses are equal; however, as
$m_\gamma = m_\psi \ll M_\phi$, and thermalization once again becomes
most important in the regime $\Trh>M_\phi$, the values taken for the
photon external mass are not numerically important.  The summed,
averaged amplitude for $\bar{\psi}\psi\leftrightarrow \gamma\gamma$
scattering is
\be
|\widebar{\mathcal{M}}(s)|^2 = \frac{m_{\psi}^2}{16\Lambda_a^2 \Lambda^{2}_b}
\left(1-\frac{4m^2_{\gamma}}{s}\right) \frac{s^3}{(s-M_{\phi}^2)^2 +
  M_{\phi}^2(\Gamma_\phi^0)^2} .
\ee
The full expression for the energy transfer rate for this case is
derived in appendix~\ref{sec:fermion-gauge}, and plotted in
figure \ref{fig:gauge-fermion} (see also the right panel of figure 
\ref{fig:eg9}).  We use the approximate analytic expression of
eq.\ \eqref{eqn:fermion-gaugexferQ} in our numerical scans.

\begin{figure}[t]
\begin{center}
\includegraphics[width = \textwidth]{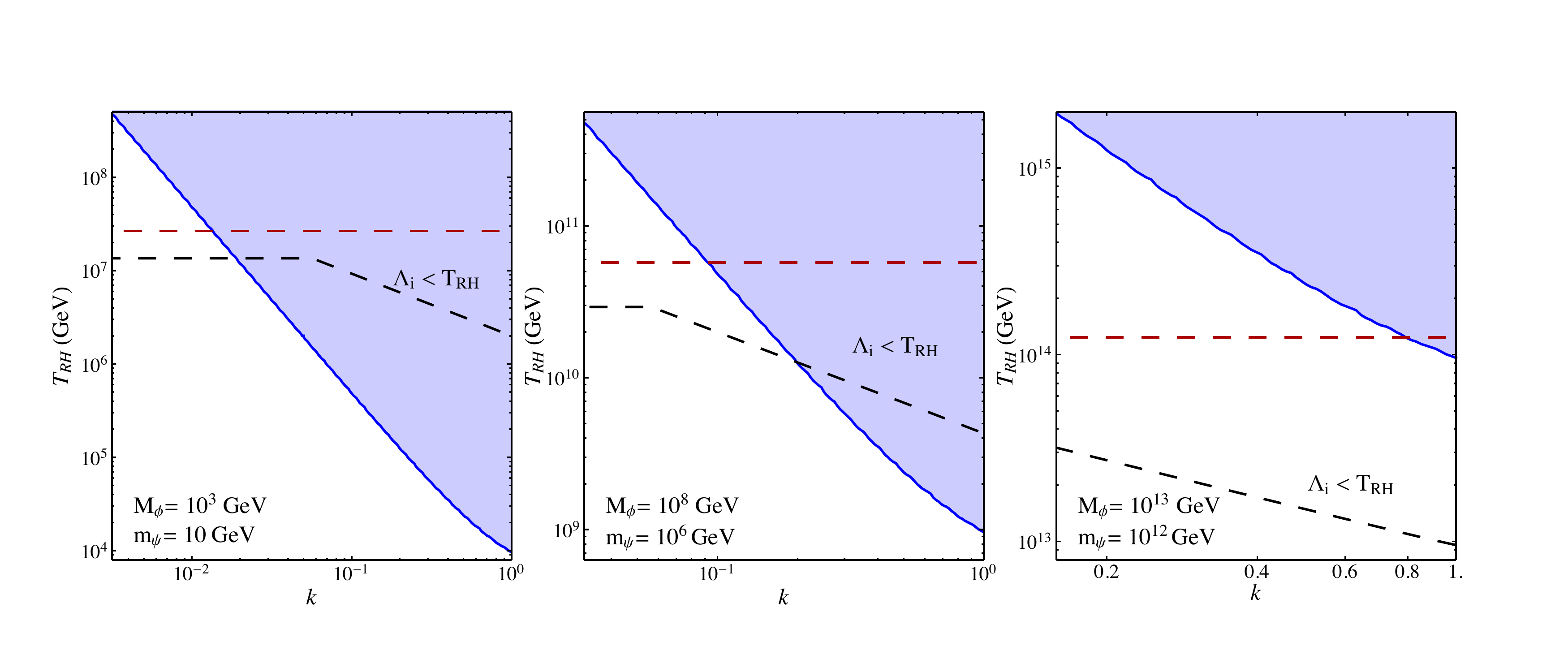}
\caption{Parameter region where thermalization can be achieved for the
  fermion-gauge model, for $M_\phi = 10^3$ GeV (left), $10^8$ GeV
  (center), and $10^{13}$ GeV (right).  Above the black dashed line,
  the EFT description of the inflaton couplings breaks down, while
  above the red dashed line preheating is always important.}
\label{fig:eg8}
\end{center}
\end{figure}

In the right panel of figure  \ref{fig:eg9} we show the evolution of the
energy transfer rate and the Hubble rate for this model. Figure 
\ref{fig:eg8} shows the regions of parameter space where the energy
transfer rate exceeds the Hubble rate for a given inflaton mass.  As
in the gauge-gauge case, the portion of thermalizing parameter space
where the EFT is reliable lies in the regime where broad resonance
preheating can be but is not necessarily realized.

\section{Discussion and Conclusions}
\label{sec:conclusions}

Asymmetric reheating is one of the minimal cosmological origins for a
dark sector that was never in thermal equilibrium with the SM, and as
such is well-motivated by a broad and growing class of DM theories
that contain multiple stable dark states.  Such theories already
frequently have trouble accommodating CMB measurements of the
effective number of neutrinos, $\Neff$, unless the dark sector is at
lower temperatures relative to the SM than are possible if the two
sectors were ever in thermal equilibrium at any stage.  As future CMB
experiments envision an order-of-magnitude improvement over Planck's
current sensitivity to $\Neff$, continued SM-like measurements of
$\Neff$ will go from restrictive to prohibitive for once-equilibrated
dark sectors containing dark radiation, as we have sketched in
section \ref{sec:deltaneff}.  More generally, the assumption that a dark
sector has thermalized with the SM is actually a fairly strong
condition on the leading interaction(s) between the two sectors, and
from this standpoint dark sectors that never thermalize with the SM
represent a generic scenario for the origin of dark matter.

Here we have systematically studied minimal models of asymmetric
reheating, demonstrating that the inevitable resulting inflaton (or
more generally `reheaton') exchange can itself thermalize the two
sectors, wiping out any would-be temperature asymmetry.  Thus
realizing a sufficiently large temperature asymmetry between two
sectors either imposes new restrictions on possible inflaton masses,
coupling structures, and reheat temperatures, or requires a
non-minimal cosmology.

We typically find that thermalization becomes important when $\Trh >
M_\phi$.  The case where reheating is dominated by trilinear inflaton
couplings to scalars in both sectors is an interesting exception, as
in this case thermalization can occur in the infrared.  In most cases,
however, thermalization is driven by potentially resonant
inflaton-mediated scattering at temperatures $\sim \Trh$.  This region
of parameter space, with $T \gtrsim M_\phi$, affords a rich variety of
interesting and model-dependent phenomena, including nonperturbative
particle production \cite{Traschen:1990sw,
  Kofman:1994rk,Kofman:1997yn} and thermal effects such as thermal
blocking and Landau damping \cite{Kolb:2003ke, Yokoyama:2005dv,
  Drewes:2010pf, Drewes:2013iaa}, and as such has been a focus of
exploration in detailed studies of reheating.  We have argued that we
generically expect both incomplete internal thermal equilibration and
nonperturbative particle creation in the form of preheating to make
inflation-mediated energy transfer between sectors more, rather than
less, important.

Our study generally serves as the first systematic study of
multi-sector reheating, and as such offers many areas for future
study. Chief amongst these is to extend our calculations to
quantitatively establish the temperature asymmetry that follows from a
given set of zero-temperature inflaton partial widths.  A full
exploration of two-sector thermalization when collective effects in
one or both radiation baths can no longer be largely neglected also
represents an obvious avenue of investigation for future
investigation, as does the question of scattering in sectors that have
not attained internal kinetic or thermal equilibrium.


\acknowledgments We thank R.~Allahverdi, M.~Amin, M.~Arvanitaki,
G.~Baym, J.~Evans, J.~Filippini, S.~Watson and especially G.~Elor for
discussions.  YC is supported by Perimeter Institute for Theoretical
Physics, which is supported by the Government of Canada through
Industry Canada and by the Province of Ontario through the Ministry of
Research and Innovation. YC is also supported in part by the Maryland
Center for Fundamental Physics. We thank the Aspen Center for Physics
for its hospitality and support through National Science Foundation
Grant No. PHY-1066293.  JS additionally thanks the Perimeter Institute
for its hospitality during the completion of this work.

\appendix

\section{Collision integrals}
\label{App:Exfer}

In this appendix, we provide a detailed derivation of our evaluation
of the collision term in the Boltzmann equation. Our derivation draws
on the derivation presented in the appendix of
ref.~\cite{Hannestad:1995rs}, which in turn is based on the earlier
work of ref.~\cite{YuehBuchler}. In this work, we have extended the
previous results by giving general expressions for the thermally
averaged cross-sections in terms of the Mandelstam variables. Further,
we derive some (to our knowledge) new analytic results in the limit
that all particles have the same mass.

\subsection{The $2\to2$ collision operator}

In this work, we are interested in processes in which two standard
model particles scatter with two hidden sector particles via inflaton
exchange. Ultimately, we are interested in the rate at which these
scatterings transfer energy between sectors; however, this rate is
closely related to the collision operator describing scattering.

We want to perform the integral\footnote{Here and in what follows,
  4-momenta are denoted by $\p$, while the magnitude of the 3-momenta are denoted 
  by $|\vec{p}| = \pp$.}
\begin{align}
\int\frac{\d^3\pp_1}{E_1(2\pi)^3}C_{\rm coll}[f] = \prod_{i = 1}^4 \int\frac{g_i\d^3 \pp_i}{2 E_i(2\pi)^3} \Lambda(f_1, f_2, f_3, f_4)\,S\,|\mathcal{M}(12\to 34)|^2(2\pi)^4\delta^{4}(\p_1+\p_2-\p_3-\p_4), 
\end{align}
where
\begin{align}
 \Lambda(f_1, f_2, f_3, f_4) = (1\pm f_1)(1\pm f_2)f_3f_4 - (1\pm f_3)(1\pm f_4)f_1f_2,
\end{align}
$|\mathcal{M}(12\to 34)|^2$ is the matrix element of the process in
question and $S$ is a possible symmetrization factor to account for identical
particles in the initial and/or final state.   For notational
simplicity, in our derivations here we will absorb the $g_i$ into an
effective symmetrization factor $\hat S$.  The plus sign here is for
bosons, while the minus sign is for fermions, accounting for the
effects of Bose enhancement and Pauli blocking, respectively. We
assume local thermal equilibrium and neglect chemical potentials, so
that the distribution functions take their equilibrium forms
\begin{align}
f_i = \frac{1}{\exp\(E_i /T\)+ \bose},
\end{align}
where $\bose = \mp 1$ for bosons and fermions, respectively, and
$\bose = 0$ for classical particles. In the limit of equilibrium, the full collision
integral vanishes.  We are interested in the rate at which
collisions occur relative to the Hubble rate. We thus consider only
the forward direction, which is equivalent to the product of the
thermally averaged cross-section and the number densities of species 1
and 2,
\begin{align}
n_1 n_2 \langle \sigma v \rangle  = \prod_{i = 1}^4 \int\frac{ \d^3
  \pp_i}{2 E_i(2\pi)^3} \hat S (1\pm f_3)(1\pm f_4)f_1f_2
|\mathcal{M}(12\to 34)|^2(2\pi)^4\delta^{4}(\p_1+\p_2-\p_3-\p_4). 
\end{align}
We now need to perform the integrations. 

We begin by writing the integral of the Lorentz-invariant phase-space
of particle 4 as
\begin{align}
\frac{\d^{3}\pp_4}{2 E_4} = \d^4\p_4 \delta(\p_4^2 - m_4^2)\Theta(\p^0_4).
\end{align}
The integral over $\d^4\p_4$ is then done using the overall
energy-conserving delta function $\delta^4(\p_1+\p_2-\p_3-\p_4)$. This
then fixes
\begin{align}
\p_4^2 = t+s+u - \p_1^2 - \p_2^2-\p_3^2,
\end{align}
where we have introduced $s$, $t$ and $u$, the standard Mandelstam
variables. The integration measures for the 3-momenta, $\vec{p}_2$ and
$\vec{p}_{3}$, are written
\begin{align}
\d^{3}\pp_2 =\pp_2^2 \d\pp_2 \d\cos\alpha\, \d\beta, \quad d^{3}\pp_3 = \pp_3^2 d\pp_3 \d\cos\theta \d\mu,
\end{align}
where the angles are defined as
\begin{align}
\cos\alpha = & \frac{{\bf p}_1\cdot{\bf p}_2}{\pp_1\pp_2}, \quad \cos\theta =  \frac{{\bf p}_1\cdot{\bf p}_3}{\pp_1 \pp_3},\\
\cos\alpha' = & \frac{{\bf p}_2\cdot{\bf p}_3}{\pp_2 \pp_3} = \cos\alpha\cos\theta + \sin\alpha\sin\theta\cos\beta.
\end{align}
We work with $s$, $t$ and $u$ instead of the angles $\alpha$, $\theta$
and $\beta$, in terms of which
\begin{align}
\label{eqn:cosacosth}
\cos\alpha =  & \frac{(E_1 + E_2)^2 -s -\pp_1^2
  -\pp_2^2}{2\pp_1\pp_2}, \quad \cos\theta = \frac{t - (E_1 - E_3)^2
  +\pp_1^2+\pp_3^2}{2\pp_1 \pp_3},\\
\cos\beta = & \frac{u - (E_2 - E_3)^2 +\pp_2^2 + \pp_3^2}{2\pp_2 \pp_3 \sin\theta\sin\alpha} - \cot\alpha\cot\theta.
\end{align}
The integration measure for $t$ and $s$ is straightforward to obtain.
To integrate over $u$ instead of $\beta$, we need to solve for
$\sin\beta$. To this end we note
\begin{align}
\sin\theta\sin\alpha\sin\beta = & \pm \sqrt{\sin^2\theta\sin^2\alpha-\( \frac{u - (E_2 - E_3)^2 +\pp_2^2 + \pp_3^2}{2\pp_2 \pp_3} - \cos\alpha\cos\theta\)^2},
\end{align}
and observe that the integration is restricted to regions where the
argument of the square root is positive (this is equivalent to
$\cos^2\beta \leq 1$). The integration measure for $\pp_2$ and $\pp_3$
is written
\begin{align}
\d^{3}\pp_2\d^{3}\pp_3 = \frac{1}{8}\frac{\d\pp_2 \d \pp_3}{\pp_1^2} d\mu \frac{\d s\, \d t\,\d u}{\sqrt{g(s, t, u)}}\Theta(g(s, t, u)).
\end{align}
Explicitly,
\begin{align}
g(s, t, u) = 1-\cos^2\alpha-\cos^2\theta+ \frac{u - (E_2 - E_3)^2 +\pp_2^2 + \pp_3^2}{\pp_2 \pp_3} \cos\alpha\cos\theta-\( \frac{u - (E_2 - E_3)^2 +\pp_2^2 + \pp_3^2}{2\pp_2 \pp_3}\)^2,
\end{align}
where from eq.\ \eqref{eqn:cosacosth}, $\cos\alpha$ and $\cos\theta$
are considered functions of $s$ and $t$, respectively. There is no
dependence on $\mu$, and the integral over this variable can be
trivially performed to give a factor of $2\pi$. We can combine this
with the rest of the expression, finding
\begin{align}\label{eqn:Colint}
\int\frac{\d^3\pp_1}{E_1(2\pi)^3}C_{\rm coll}[f] = \frac{1}{16(2\pi)^6}
\int \frac{\d\pp_1 \d\pp_2 \d \pp_3}{E_1 E_2 E_3}  \frac{\d s \d t\d
  u}{\sqrt{g}}\Theta(g) \Lambda(f_1, f_2, f_3, f_4)\hat S|\mathcal{M}|^2
\delta\(s+t+u - \sum_i m^2_i\), 
\end{align}

For a completely general matrix element, this is as far as we can go
in full generality. We can make some more progress if we assume, as
appropriate for the interactions important in this work, that the
matrix element is a function only of either $s$, $t$, or $u$.

For each variable, note that $g(s, t, u)$ is a quadratic function of
$s$, $t$, and $u$.  After the integration over the delta function is
performed, it is a quadratic function of only two of $s$, $t$ or $u$,
and we can make use of the fundamental identity
\cite{Hannestad:1995rs,YuehBuchler}
\begin{align}\label{eqn:identity}
\int \frac{\d x}{\sqrt{a x^2 + b x + c}}\Theta(a x^2 + b x + c) = \frac{\pi}{\sqrt{-a}} \Theta(b^2 - 4 ac),
\end{align}
where $x$ is any of $s$, $t$ or $u$, and $a$, $b$ and $c$ are
(quadratic) functions of the remaining Mandelstam variable. In this
work we are interested primarily in energy transfer between
sectors. As discussed above in section \ref{sec:Exfer}, energy transfer is
dominated by $s$-channel processes in the regimes of interest. We thus
focus on these.  Similar results for $u$ and $t$ channel amplitudes
can be readily obtained following the same steps.

\subsection{$s$-channel scattering}

Assuming that the matrix element depends only on $s$, we integrate
over $u$ using the delta function, and make use of eq.\ 
\eqref{eqn:identity} to integrate over $t$. In this case, the
parameter $a$ on the right hand side of eq.\  \eqref{eqn:identity} is
\begin{align}
a = -\frac{((E_1 + E_2)^2 - s)}{4\pp_1^2\pp_2^2 \pp_3^2}.
\end{align}
The Heaviside function, $\Theta(b^2 - 4 ac)$, restricts the integration to regions where 
\begin{align}\label{eqn:slimits}
A = b^2 - 4 ac = \frac{\(\tilde{s} + (\pp_1+\pp_2)^2\)\(\tilde{s}+(\pp_3+\pp_4)^2\)\(\tilde{s}+ (\pp_1-\pp_2)^2\)\(\tilde{s}+(\pp_3-\pp_4)^2\) }{(4\pp_1^2\pp_2^2 \pp_3^2)^2}> 0.
\end{align}
In this expression $\tilde{s} = s-(E_1+E_2)^2$ and
\begin{align}
\pp_4 = \sqrt{(E_1+E_2 - E_3)^2-m_4^2}.
\end{align}
After switching integration variables to $E_1$, $E_2$ and $E_3$, eq.\ \eqref{eqn:Colint} can be written\begin{align}
\int\frac{\d^3\pp_1}{E_1 (2\pi)^3}C_{coll}[f] = \frac{\pi}{8(2\pi)^6}
\int  \frac{ \d E_1 \d E_2 \d E_3 \d s }{\sqrt{((E_1 + E_2)^2 -
    s)}}\Theta(A) \Lambda(f_1, f_2, f_3, f_4)\hat S|\mathcal{M} (12\to 34)|^2.
\end{align}

To make further progress, we need to find the region of integration
where eq.\ \eqref{eqn:slimits} holds.  To this end, recall that by
definition (see eq.\ \eqref{eqn:cosacosth})
\begin{align}
-(\pp_1 +\pp_2)^2 < s-(E_1+E_2)^2 < -(\pp_1 -\pp_2)^2,
\end{align}
so in order to have a non-zero integration region, $A>0$, we need to find the regions where both
\begin{align}
0 < s-(E_1+E_2)^2 + (\pp_3+\pp_4)^2,
\text{ and }
 s-(E_1+E_2)^2 + (\pp_3-\pp_4)^2 < 0
\end{align}
hold. 

There are then four cases which can lead to non-zero integration regions with $A>0$ 
\begin{enumerate}
\item $0 < (\pp_3+\pp_4)^2-(\pp_1 +\pp_2)^2$ and $0 > (\pp_3-\pp_4)^2-(\pp_1 -\pp_2)^2$ with $-(\pp_1 +\pp_2)^2 < s-(E_1+E_2)^2 < -(\pp_1 -\pp_2)^2$.

\item $0 > (\pp_3+\pp_4)^2-(\pp_1 +\pp_2)^2$ and  $0 < (\pp_3-\pp_4)^2-(\pp_1 -\pp_2)^2$ with $-(\pp_3 + \pp_4)^2 < s-(E_1+E_2)^2 < -(\pp_3 - \pp_4)^2$.

\item $0 < (\pp_3+\pp_4)^2-(\pp_1 +\pp_2)^2$ and $0 < (\pp_3-\pp_4)^2-(\pp_1 -\pp_2)^2$ with $-(\pp_1 +\pp_2)^2 < s-(E_1+E_2)^2 < -(\pp_3 - \pp_4)^2$, \emph{and} $\pp_1+\pp_2 > \pp_3-\pp_4$.

\item $0 > (\pp_3+\pp_4)^2-(\pp_1 +\pp_2)^2$ and $0 > (\pp_3-\pp_4)^2-(\pp_1 -\pp_2)^2$ with $-(\pp_3 + \pp_4)^2 < s-(E_1+E_2)^2 < -(\pp_1 -\pp_2)^2$ \emph{and} $\pp_3+\pp_4 >\pp_1-\pp_2$.

\end{enumerate}
The limits in eq.\ \eqref{eqn:slimits} can be used to restrict
the integration over $E_3$. In general, the radicals that appear when the three momentum,
$\pp$, is replaced by the corresponding energy $E$ in these conditions make them
difficult, if not intractible to reduce analytically. However, if we
assume all particles have identical masses, some progress can be made.

Assuming all four particles have identical masses, $m$, we find that
only the first two regions 1.\ \& 2.\ above contribute and lead to the
following restrictions. 

\begin{itemize}
\item From the first case above, the integration
limits are
\begin{align}\nn
E_2 < E_3 & < E_1,  \text{ or } E_1< E_3 <E_2,\\
|E_1-E_2| & < \sqrt{\((E_1+E_2)^2-s\)\(1-\frac{4 m^2}{s}\)}, \quad E_1+E_2 > \sqrt{s},\quad  s> 4m^2.
\end{align}
\item For the second case, we find for $E_2 < E_1$
\begin{align}
 E_1 < E_3 & < E_1+E_2 - m,  \text{ or } m < E_3 < E_2,
\end{align}
and for $E_1< E_2$
\begin{align}
E_2 < E_3 <E_1+E_2-m, \text{ or } m < E_3 < E_1,
\end{align}
and
\begin{align}
|E_3-E_4| & < \sqrt{((E_1+E_2)^2-s)\(1-\frac{4 m^2}{s}\)}, \quad E_1+E_2 > \sqrt{s},\quad  s> 4m^2,
\end{align}
where $E_4 = E_1 + E_2 - E_3$. 
\end{itemize}
If we further assume that particles $1$
and $2$ obey the same statistics, as do particles $3$ and $4$, we find
\begin{align}\nn \label{eqn:TAXsection}
n_1 n_2 \langle \sigma v \rangle =  & \frac{\pi}{(2\pi)^6}\frac{1}{8}\int_{4m^2}^{\infty} \hat S |\mathcal{M}(s)|^2 \d s \int_{\sqrt{s}}^{\infty} \frac{\d E_+}{\sqrt{E_+^2 - s}}\Bigg[ \int_0^{|E_-|_{\rm max}}\!\!\!\!\!\d E_- \int_{\frac{1}{2}(E_++E_-)}^{\frac{1}{2}(E_+-E_-)} \!\!\!\!\d E_3 (1\pm f_3)(1\pm f_4)f_1 f_2 \\  &  \hskip 3.5 cm+ \int_0^{|\tilde{E}_-|_{\rm max}}\d\tilde{E}_- \int_{\frac{1}{2}(E_++\tilde{E}_-)}^{\frac{1}{2}(E_+-\tilde{E}_-)} \d E_2 (1\pm f_3)(1\pm f_4)f_1 f_2 \Bigg],
\end{align}
where $E_+ = E_1 + E_2$ and $E_- = E_1 - E_2$, and $\tilde{E}_- =
E_3-E_4$.

Next, we can integrate over the distributions. The integrals over
$E_3$ and $E_2$ are easily done, yielding
\begin{align}\label{eqn:intoverE3}
\int_{\frac{1}{2}(E_++E_-)}^{\frac{1}{2}(E_+-E_-)} \!\!\!\!\!\d E_3 (1-\bose_a f_3)(1- \bose_a f_4)
= & \frac{ T}{1 - \bose_a^2 e^{-\frac{E_+}{T}}} \(\ln\[\frac{\bose_a +  e^{\frac{E_+-E_-}{2T}}}{1 + \bose_a e^{-\frac{E_+ + E_-}{2T}}}\]-\ln\[\frac{\bose_a+ e^{\frac{E_++E_-}{2T}}}{1 + \bose_a e^{\frac{E_-- E_+}{2T}}}\]\),
\end{align}
and
\begin{align}\label{eqn:intoverE2}
\int_{\frac{1}{2}(E_++\tilde{E}_-)}^{\frac{1}{2}(E_+-\tilde{E}_-)} \d E_2 f_1 f_2
= & \frac{ T}{1 - \bose_b^2 e^{-\frac{E_+}{T}}} \(\ln\[\frac{\bose_b +  e^{\frac{E_+-\tilde{E}_-}{2T}}}{1 + \bose_b e^{-\frac{E_+ + \tilde{E}_-}{2T}}}\]-\ln\[\frac{\bose_b+ e^{\frac{E_++\tilde{E}_-}{2T}}}{1 + \bose_b e^{\frac{E_-- \tilde{E}_+}{2T}}}\]\),
\end{align}
where $\bose_a, \bose_b = 1$ for Fermi-Dirac statistics,
$\bose_a,\bose_b = -1$ for Bose-Einstein and $\bose_a,\bose_b = 0$ for
Maxwell-Boltzmann. In the case where particles 3 and 4 obey different
statistics to particles 1 and 2, the remaining three integrations
(after inserting Eqns.\ \eqref{eqn:intoverE3} and
\eqref{eqn:intoverE2} into eq.\ \eqref{eqn:TAXsection}) need to be
performed numerically. If all particles obey the same statistics,
$\bose_a = \bose_b$, we can analytically integrate over $E_-$ and
$\tilde{E}_-$ as well, in which case we arrive at
\begin{align}\label{eqn:finres}
n_1 n_2 \langle \sigma v \rangle = & \frac{\pi}{(2\pi)^6}\frac{T^2 }{2}\int_{4m^2}^{\infty} |\mathcal{M}(s)|^2 \d s \int_{\sqrt{s}}^{\infty} \frac{\d E_+}{\sqrt{E_+^2 - s}}\frac{  e^{-\frac{E_+}{T}}}{\(1-\bose_a^2 e^{-\frac{E_+}{T}}\)^2}\log
   ^2\left(\frac{e^{\frac{E_{+}}{2T}}+\bose_ae^{\frac{\sqrt{ (E_+^2-s)\(1 -\frac{4 m^2}{s}\)}}{2T}}}{e^{\frac{\sqrt{ (E_+^2-s)\(1 -\frac{4 m^2}{s}\)}+E_{+}}{2T}}+\bose_a}\right).
\end{align}
In general, these final two integrations must be performed
numerically. Similarly, the energy transfer rate is given by the
integral
\begin{align}\label{eqn:finresExfer}
n_1 n_2 \langle \sigma v (E_1&+E_2)\rangle\\ \nn = & \frac{\pi}{(2\pi)^6}\frac{T^2 }{2}\int_{4m^2}^{\infty} |\mathcal{M}(s)|^2 \d s \int_{\sqrt{s}}^{\infty} \frac{\d E_+}{\sqrt{E_+^2 - s}}\frac{   E_+ e^{-\frac{E_+}{T}}}{\(1-\bose_a^2 e^{-\frac{E_+}{T}}\)^2}\log
   ^2\left(\frac{ e^{\frac{E_{+}}{2T}}+\bose_ae^{\frac{\sqrt{ (E_+^2-s)\(1 -\frac{4 m^2}{s}\)}}{2T}}}{e^{\frac{\sqrt{ (E_+^2-s)\(1 -\frac{4 m^2}{s}\)}+E_{+}}{2T}}+\bose_a}\right).
\end{align}

\subsubsection{Maxwell-Boltzmann limit}

In the limit that all particles obey Maxwell-Boltzmann statistics,
$\bose_a = 0$, $\bose_b = 0$, we can evaluate the integral over $E_+$
analytically, and recover the known results of Gelmini and
Gondolo \cite{Gondolo:1990dk}.

In the Maxwell-Boltzmann limit, eq.\ \eqref{eqn:finres} becomes
\begin{align}
n_1n_2\langle \sigma v \rangle= & \frac{\pi}{(2\pi)^6}\frac{1}{8}\int_{4m^2}^{\infty}\d s \(1 -\frac{4 m^2}{s}\) \hat S |\mathcal{M}(s)|^2 \int_{\sqrt{s}}^{\infty} \d E_+ \sqrt{E_+^2 - s}\;e^{-\frac{E_+}{T}}\\
= & \frac{\pi}{(2\pi)^6}\frac{T}{8}\int_{4m^2}^{\infty}\d s\(s -4 m^2\) \hat S \frac{|\mathcal{M}(s)|^2 }{s}\sqrt{s}K_{1}\(\frac{\sqrt{s}}{T}\),
\end{align}
where here and below the $K_n(x)$ are modified Bessel functions of the
second kind of order $n$. In the center of mass frame, 
\begin{align}
\frac{d\sigma}{d\Omega} = \frac{|\mathcal{M}|^2}{64\pi^2 s}
\end{align}
so that if $\mathcal{M}$ depends only on $s$,
\begin{align}
\sigma = \frac{|\mathcal{M}|^2}{16\pi s}.
\end{align}
Making use of the Maxwell-Boltzmann limit result
\begin{align}
n_1 n_2 = \frac{1}{(2\pi)^6}\[4\pi m^2 T K_2\(\frac{m}{T}\)\]^2
\end{align}
we find the thermally averaged cross-section in the Maxwell-Boltzmann
limit
\begin{align}
\langle \sigma v \rangle =\frac{1}{8 m^4 T K_2\(\frac{m}{T}\)^2}\int_{4m^2}^{\infty}ds\(s -4 m^2\)\sigma \sqrt{s}K_{1}\(\frac{\sqrt{s}}{T}\) 
\end{align}
in agreement with the result of Gelmini and Gondolo  \cite{Gondolo:1990dk}. We write this as
\begin{align}\label{eqn:xsection}
n_1 n_2 \langle \sigma v \rangle =  \frac{2\pi^2}{(2\pi)^3}\frac{ T}{16\pi}\int_{4m^2}^{\infty}\d s\(s -4 m^2\)\hat S \frac{|\mathcal{M}(s)|^2}{s} \sqrt{s}K_{1}\(\frac{\sqrt{s}}{T}\) 
\end{align}
A similar integral form can be found for the energy transfer rate
\begin{align}\label{eqn:Etrans}\nn
n_1 n_2 \langle E_+ \sigma v \rangle = & \frac{\pi}{(2\pi)^6}\frac{1}{8}\int_{4m^2}^{\infty}\d s \(1 -\frac{4 m^2}{s}\)\hat S |\mathcal{M}(s)|^2 \int_{\sqrt{s}}^{\infty} \d E_+ E_+ \sqrt{E_+^2 - s}\; e^{-\frac{E_+}{T}}\\
= & \frac{2\pi^2}{(2\pi)^6}\frac{T}{16\pi}\int_{4m^2}^{\infty}\d s\(s -4 m^2\) \hat S |\mathcal{M}(s)|^2 K_{2}\(\frac{\sqrt{s}}{T}\).
\end{align}

\section{Energy transfer rates}
\label{App:specifics}

In this section we consider specific simple models as discussed in
section \ref{sec:examplesReh}, and develop some approximations that allow
for analytic integration of the thermally averaged cross-section, as
well as the energy transfer rate, in the Maxwell-Boltzmann limit. Away
from the resonance, only small corrections are required to obtain good
agreement with the full numerical evaluation including full quantum
statistics. 

For numerical efficiency, we make use of the approximations derived
below in our scans over parameter space in
section \ref{sec:examplesReh}. In all cases, we do not expect that the
poorer agreement of our approximation with the exact result near $T
\sim M_{\phi}$ will significantly affect the results of our
scans. This can be seen from figure  \ref{fig:eg1} and figure 
\ref{fig:eg9}. Notice that in each of the limiting cases illustrated
in these figures, $H(T)$ intersects $\Gamma^E(T)$ at the temperature
where $\Gamma^E(T)$ is at a local maximum, which occurs in the region
$T < M_{\phi}$\footnote{This behavior is straightforward to
  understand. Near the resonance, $\Gamma^E$ takes the form $x^{-3}
  K_{2}(1/x)$, with $x = T/M_{\phi}$. The function $K_2(x)$ can be
  approximated near $x\sim 0$ as $x^{1/2}\exp(1/x)$. The function
  $x^{-5/2} \exp(1/x)$ peaks at $x_{\rm peak} \sim 2/5 $, and thus the
  local maximum due to the resonance occurs at $T = 2M_{\phi}/5$,
  somewhat below $T \sim M_{\phi}$.} where our approximations are
excellent. Further, note that whenever our approximations mis-estimate
$\Gamma^{E}$, the combination of the shape of resonant feature in
$\Gamma^E$ and our scan strategy that looks at all $T < \Trh$ means
that we will not misattribute thermalized regions as unthermalized or
vise-versa. However, if one were only interested in the condition
$\Gamma^E(\Trh) > H(\Trh)$ at a single fixed $\Trh$, then these
approximations would be inadequate.

All numerical result in this section were obtained using
the publicly available
Cubature,\footnote{http://ab-initio.mit.edu/wiki/index.php/Cubature}
and Cuba\footnote{http://www.feynarts.de/cuba/} \cite{Hahn:2004fe}
libraries for C. 

\subsection{Scalar boson scattering}
\label{sec:scaltrilin}

\begin{figure}[h!]
\centering
\includegraphics[scale = 0.4]{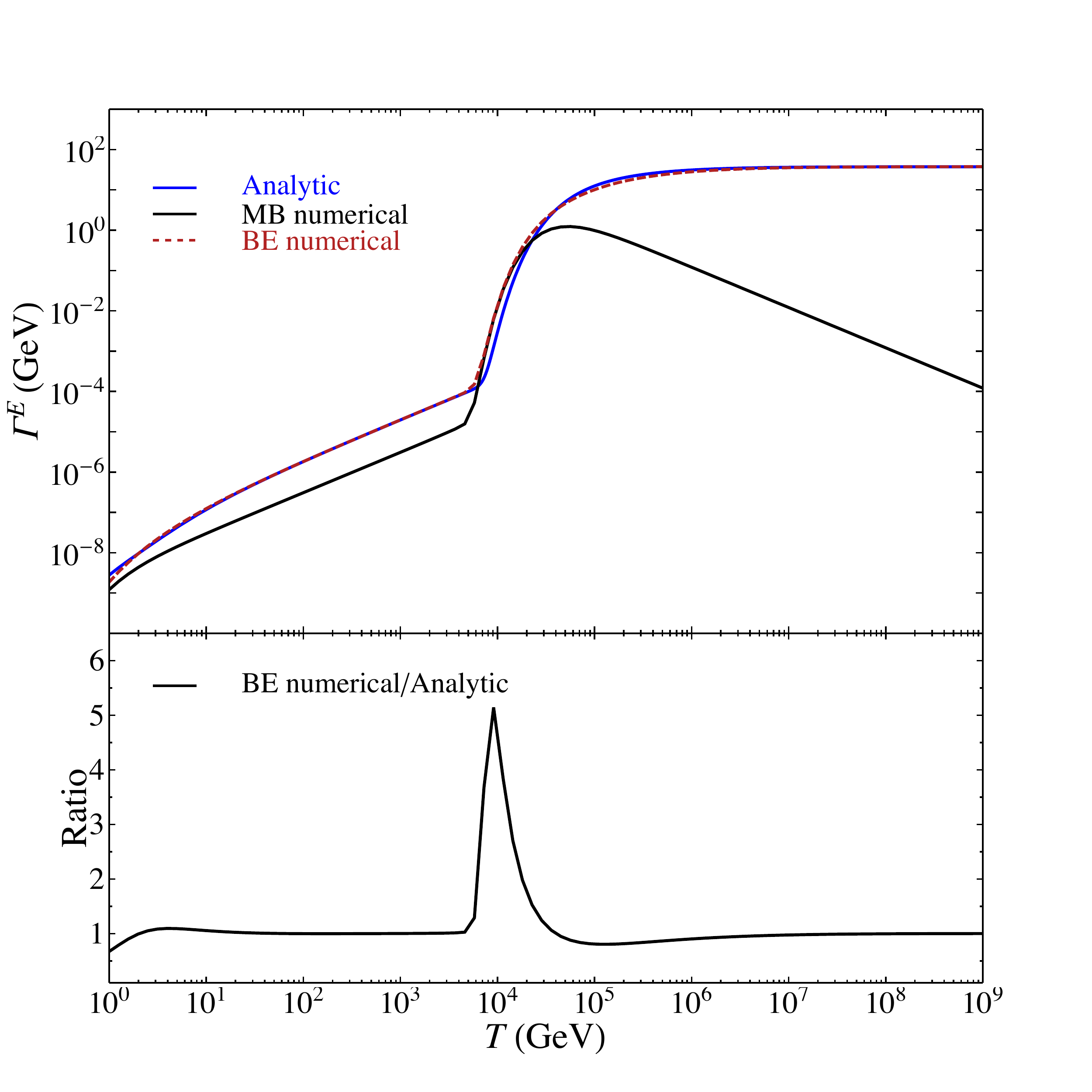}
\caption{Upper panel: We show the energy transfer rate for a scalar
  trilinear interaction for $m = m_S = 1$ GeV, $M_{\phi} = 10^5$ GeV,
  $w = 10^{-4}$. Plotted is the numerical result using
  Maxwell-Boltzmann statistics (black curve), the numerical result
  using Bose-Einsten statistics (red curve) and the analytic
  approximation of eq.\ \eqref{eqn:scalartrilinapprox} (blue
  curve). Lower panel: In the lower panel we show the ratio of the
  numerical result evaluated with Bose-Einstein statistics to the analytic
  approximation.}\label{fig:scalartrilinear}
\end{figure}

We first consider the scattering of real scalars, $S_a$, $S_b$, via a
trilinear interaction with a third scalar $\phi$, as described in
eqs.\ \eqref{eq:scalarLag1}--\eqref{eq:scalarLag3}. 
To a very good approximation, the amplitude for this process can be
approximated by treating the resonance as infinitely sharp and taking
the matrix element to be simply a constant below the resonance and
vanishing above,
\begin{align}
|\widebar{\mathcal{M}}(s)|^2 = \frac{\mu^2_a \mu^2_b}{ M_{\phi}^4}\Theta(M_{\phi} - \sqrt{s}) + \mu^2_a \mu^2_b\frac{\pi}{M_\phi \Gamma_{\phi}}\delta(s-M_{\phi}).
\end{align}
Somewhat surprisingly, comparison of a direct numerical evaluation of
the thermally averaged cross-section in the Maxwell-Boltzmann limit
shows that this is an excellent approximation even when
$\Gamma_{\phi}\lesssim M_{\phi}$.

In the case where all external particles obey Maxwell-Boltzmann
statistics, it is straightforward to obtain for the thermally averaged
scattering cross-section from eq.\ \eqref{eqn:xsection}
\begin{align}\nn
n_1 n_2\langle \sigma v \rangle 
 = &\frac{\mu^2_a \mu^2_b}{4}\frac{1}{(2\pi)^6} \frac{2\pi^2}{16 \pi} \frac{T}{M_{\phi}^4}\[ 8T^2 m  K_1\( \frac{2m}{T}\) - 2T\( ( M_{\phi}^2-4m^2) K_0 \( \frac{M_{\phi}}{T}\)  + 
   2 M_{\phi} T K_1\( \frac{M_{\phi}}{T}\)\)\]\\ & +\frac{\mu^2_a \mu^2_b}{4}\frac{1}{(2\pi)^6} \frac{2\pi^2 T}{16 \pi} \frac{\pi}{M_{\phi} \Gamma_{\phi}}\(1-4\frac{m^2}{M_{\phi}^2}\) M_{\phi}K_{1}\(\frac{M_{\phi}}{T}\),
\end{align}
where the functions $K_i(x)$ are the modified Bessel
functions and $m = m_S$ is the mass of the scalar boson. Similarly, we can obtain the thermally averaged
energy-transfer rate using eq.\ \eqref{eqn:Etrans}
\begin{align}\nn
n_1 n_2 \langle \sigma v (E_1 +E_2)\rangle %
= &\frac{\mu^2_a \mu^2_b}{4}\frac{1}{(2\pi)^6} \frac{2\pi^2}{16 \pi} \frac{T}{M_{\phi}^4}\[8 m^2 T \left(2 m K_3\left(\frac{2 m}{T}\right)-T G_{1,3}^{3,0}\left(\frac{m^2}{T^2}\bigg|
\begin{array}{c}
 1 \\
 0,0,2 \\
\end{array}
\right)\right)\]\\\nn
& -\frac{\mu^2_a \mu^2_b}{4}\frac{1}{(2\pi)^6} \frac{2\pi^2}{16 \pi} \frac{T}{M_{\phi}^4}\[ 2 T \left(M_{\phi}^3 K_3\left(\frac{M_{\phi}}{T}\right)-4 m^2 T G_{1,3}^{3,0}\left(\frac{M_{\phi}^2}{4 T^2}\bigg|
\begin{array}{c}
 1 \\
 0,0,2 \\
\end{array}
\right)\right)\]\\
 &  +\frac{\mu^2_a \mu^2_b}{4}\frac{1}{(2\pi)^6} \frac{2\pi^2}{16 \pi} T \frac{\pi}{M_\phi \Gamma_\phi}  (M_{\phi}^2 - 4m^2)  K_{2}\(\frac{M_{\phi}}{T}\)
\end{align}
Here $G_{1,3}^{3,0}(x)$ is the Meijer G-function. It is the subleading
part of the expression at high $T/m$ and can be dropped in this
regime, while it changes the result by order unity at low $T/m \sim
1$.

For Bose-Einstein statistics, the energy transfer rate at eq.\ 
\eqref{eqn:finresExfer} is well-approximated by
\begin{align}\nn\label{eqn:scalartrilinapprox}
n_1 n_2 \langle (E_1+E_2)\sigma v \rangle \approx  \frac{\mu^2_a \mu^2_b }{4(2\pi)^6}\frac{2\pi^2}{16\pi}\frac{T^2}{M_{\phi}^4} \Bigg[ &A \frac{\pi^5}{8}Tm^2 \(2K_{2}\(\frac{2m}{T}\) - K_1\(\frac{2m}{T}\) - 2\frac{T}{m}K_0\(\frac{2m}{T}\)\)\Theta(M_\phi-T)\\ & +4 \pi^2 B \frac{T M_{\phi}^3}{\Gamma_{\phi}}\(\pi K_1\(\frac{M_{\phi}}{T}\) - 2 K_0\(\frac{M_{\phi}}{T}\)\)\Bigg],
%
\end{align}
where $A = 1.06$ and $B = 0.615$. As can be seen in
figure \ref{fig:scalartrilinear}, this approximation is excellent away
from the resonance region.  It fails completely for $T \sim
m$. Unfortunately, we have been unable to derive this result in a
consistent fashion by approximating the integrals of eq.\ 
\eqref{eqn:finres}.

\subsection{Fermion scattering}
\label{sec:fermifermi}

\begin{figure}[h!]
\centering
\includegraphics[scale = 0.4]{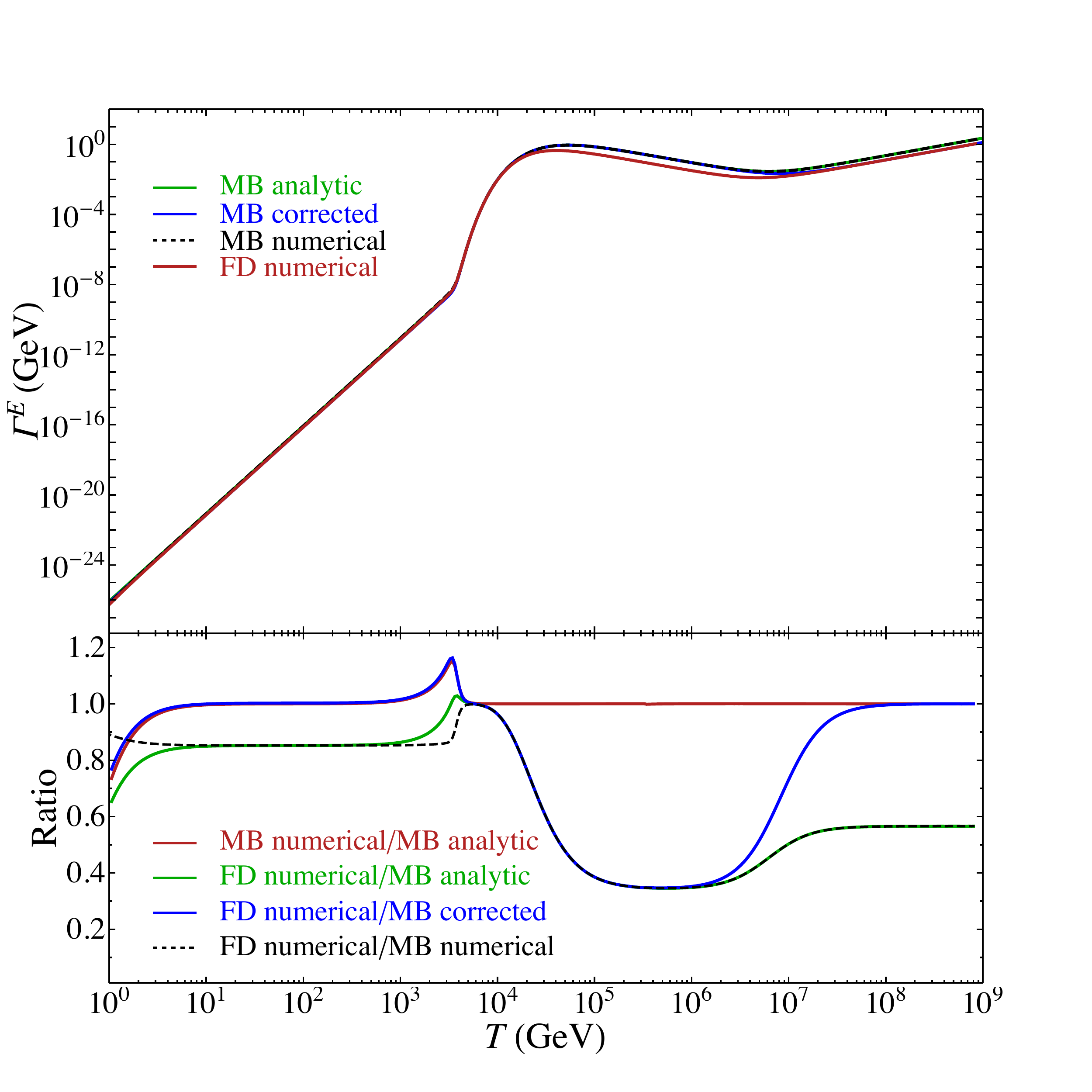}
\caption{Upper panel: We show the energy transfer rate for
  scalar-mediated fermion scattering. Parameters are chosen to be $m = m_\psi =
  1$ GeV, $M_{\phi} = 10^5$ GeV, $w = 10^{-4}$. Shown are the rates for
  Fermi-Dirac statistics (red), the numerically evaluated
  Maxwell-Boltzmann rate in black dashed and the analytic
  approximations at \eqref{eqn:fermionyukawaMB} (green) and
  \eqref{eqn:fermionyukawaFDapp} (blue). Lower panel: Black dashed
  shows the ratio of the Maxwell-Boltzmann energy transfer rate to the
  Fermi-Dirac energy transfer rate. In red, we show the ratio of the
  approximation of eq.\  \eqref{eqn:fermionyukawaMB} and the
  numerically evaluated energy transfer rate for Maxwell-Boltzmann
  statistics. In green we show the ratio the approximation
  \eqref{eqn:fermionyukawaMB} to the Fermi-Dirac transfer rate. In
  blue we show the ratio of the approximation
  \eqref{eqn:fermionyukawaFDapp} to the Fermi-Dirac transfer
  rate.}\label{fig:fermion-yukawa}
\end{figure}

We next consider the scattering of Dirac fermions $\psi_a$, $\psi_b$, via Yukawa
interactions with a scalar $\phi$, as described in
eqs.\  \eqref{eq:fermiLag1}--\eqref{eq:fermiLag3}. 
Analogous to the scalar trilinear interaction considered above in
section \ref{sec:scaltrilin}, to a good approximation, we can approximate
this matrix element using an infinitely sharp resonance with power law
(in $s$) behaviour away from the resonance region
\begin{align}\nn
\left|\widebar{\mathcal{M}}(s)\right|^2 \approx & y_a^2y_b^2 \(1-\frac{4m^2}{s}\)^2 \frac{s^2}{M_\phi^4}\Theta(M_\phi^2-s) +y_a^2y_b^2 \frac{\pi}{\Gamma_\phi M_{\phi}}\(1-\frac{4m^2}{M_{\phi}^2}\)^2M_{\phi}^4\delta(s-M_{\phi}^2)\\
& +y_1^2y_2^2\(1-\frac{4m^2}{s}\)^2  \Theta(s-M_\phi^2),
\end{align}
where $m = m_\psi$ is the fermion mass.

If we neglect quantum statistics and treat the fermions as classical
particles, we can evaluate  the energy transfer rate
\begin{align}\label{eqn:fermionyukawaMB}
n_1 n_2 \langle \sigma v(E_1 + E_2)\rangle =4 y_a^2y_b^2\frac{1}{(2\pi)^6} \frac{2\pi^2}{16\pi}T\Big[& \(1-\frac{4m^2}{M_{\phi}^2}\)^3 \frac{\pi}{\Gamma_{\phi}}M_{\phi}^5 K_2\(\frac{M_{\phi}}{T}\)\\\nn & +2^{11}3 \frac{T^8}{M_{\phi}^4} \Theta(M_{\phi}-T) +2M_{\phi}^3 T K_3\(\frac{M_{\phi}}{T}\)\Theta(T-M_{\phi}) \Big].
\end{align}
The cross-section can also be found using the same technique. In obtaining this result, as described above, we have expanded in the large $T/m$ and $T/M_{\phi}$ limits.

Away from the Maxwell-Boltzmann limit, analytic solution of the energy
transfer rate is difficult. We can, however, numerically evaluate
eq.\ \eqref{eqn:finresExfer}, finding that away from the resonance region it is well-approximated by
\begin{align}\label{eqn:fermionyukawaFDapp}
n_1 n_2 \langle \sigma v(E_1 + E_2)\rangle =4 y_a^2y_b^2 \frac{1}{(2\pi)^6} \frac{2\pi^2}{16\pi}T\Big[& \(1-\frac{4m^2}{M_{\phi}^2}\)^3 \frac{\pi}{\Gamma_{\phi}}M_{\phi}^5 K_2\(\frac{M_{\phi}}{T}\)\\\nn & +A 2^{11}3 \frac{T^8}{M_{\phi}^4} \Theta\(\frac{M_{\phi}}{10}-T\) +2BM_{\phi}^3 T K_3\(\frac{M_{\phi}}{T}\)\Theta(T-M_{\phi}) \Big],
\end{align}
where $A \approx 0.85$ and $B = \frac{2\sqrt{2}}{5}$. The results of
these approximations are shown in figure  \ref{fig:fermion-yukawa}.

\subsection{Gauge boson scattering}
\label{sec:gaugegauge}

\begin{figure}[h!]
\centering
\includegraphics[scale = 0.4]{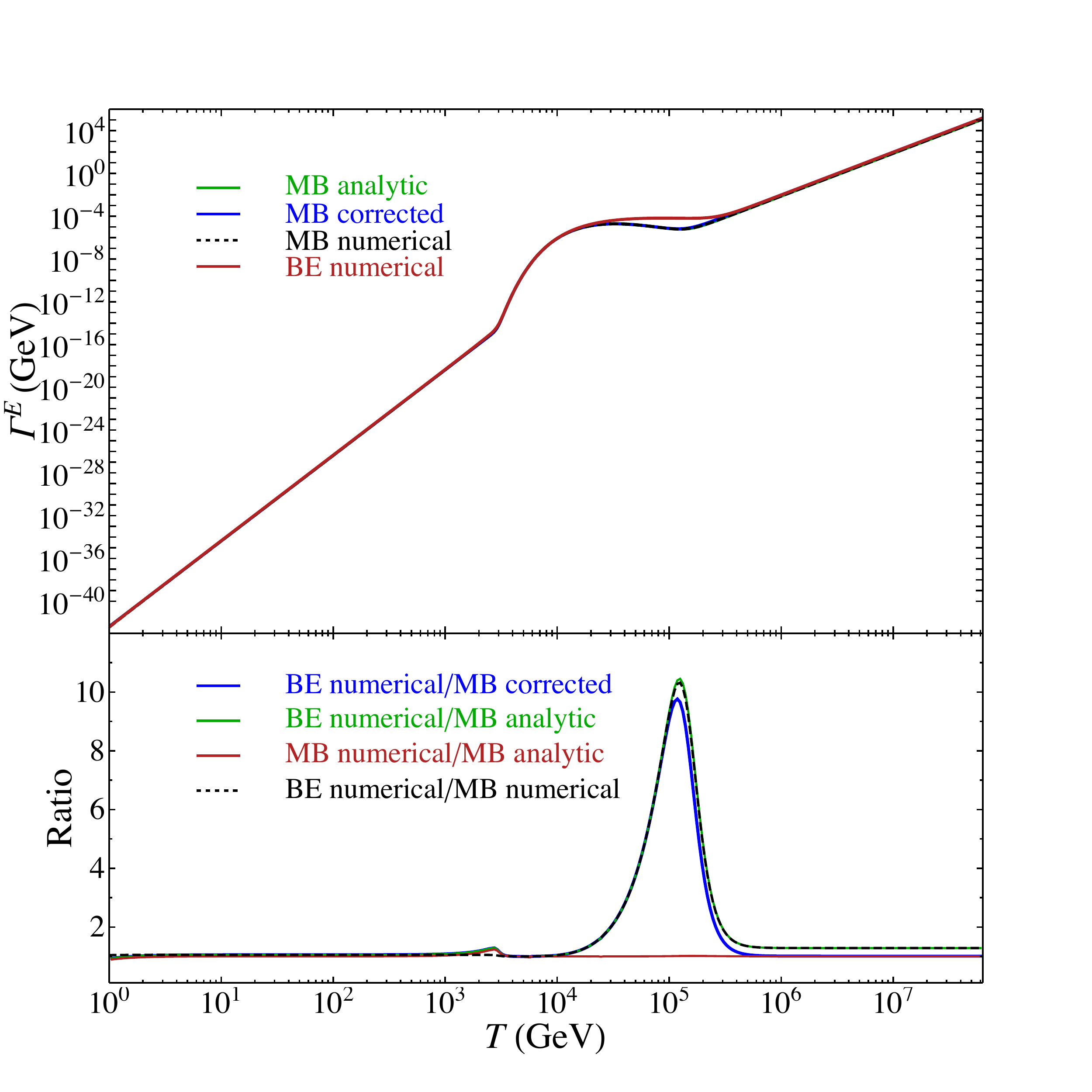}
\caption{Upper panel: We show the energy transfer rate for
  axion-mediated gauge boson scattering. Parameters are chosen to be
  $m = m_\gamma = 1$ GeV, $M_{\phi} = 10^5$ GeV, $w = 10^{-4}$. Shown
  are the rates for Bose-Einstein statistics (red), the numerically
  evaluated Maxwell-Boltzmann rate (black, dashed) and the analytic
  approximations at eq.\ \eqref{eqn:gaugembenergyxfer} (green) and eq.\
  \eqref{eqn:gaugembenergyxferBE} (blue). Lower panel: We show the
  ratio of the energy transfer rates. Black dashed shows the ratio of
  the Maxwell-Boltzmann energy transfer rate to the Bose-Einstein
  energy transfer rate. In red, we show the ratio of the approximation
  of eq.\  \eqref{eqn:gaugembenergyxfer} and the numerically evaluated
  energy transfer rate for Maxwell-Boltzmann statistics. In green we
  show the ratio of the approximation eq.\ \eqref{eqn:gaugembenergyxfer} to
  the Bose-Einstein transfer rate. In blue we show the ratio of the
  approximation eq.\ \eqref{eqn:gaugembenergyxferBE} to the Bose-Einstein
  transfer rate. The error at the resonance scales with the width,
  while the asymptotics are independent of the
  width.}\label{fig:gauge-axion}
\end{figure}

We now consider the scattering of gauge bosons via a dimension-five
interaction with a pseudoscalar $\phi$, as described in
eqs.\ \eqref{eq:gaugeLag1}--\eqref{eq:gaugeLag3}.  Analogous to the above
cases, this matrix element is well-approximated by
\begin{align}\nn
|\widebar{\mathcal{M}}(s)|^2 \approx & \frac{1}{128\Lambda_a^2\Lambda_b^2}\(1-\frac{4m^2}{s}\)^2 \frac{s^4}{M_\phi^4}\Theta(M_\phi^2-s) + \frac{\pi}{\Gamma_\phi M_{\phi}}\frac{1}{128\Lambda_a^2\Lambda_b^2}\(1-\frac{4m^2}{M_{\phi}^2}\)^2M_{\phi}^8\delta(s-M_{\phi}^2)\\
& +\frac{1}{128\Lambda_a^2\Lambda_b^2}\(1-\frac{4m^2}{s}\)^2 s^2 \Theta(s-M_\phi^2),
\end{align}
where $m = m_\gamma$ is the gauge boson mass.

In this case, if we ignore quantum statistics, all integrations are
analytically tractable and the result for the cross-section can be
obtained in terms of Meijer G-functions. However, the Meijer
G-functions are not easily implemented numerically, and do not provide much insight. For these reasons we expand our results as power series in $T/M_{\phi}$ and $T/m$.  For $m < T$ and and $M_{\phi} < T$, to an excellent approximation we can expand and
keep only the leading order behavior in powers of $T$
\begin{align}\label{eqn:gaugembxsection}\nn
n_1 n_2 \langle \sigma v \rangle =   \frac{1}{128\Lambda_a^2\Lambda_b^2}\frac{1}{(2\pi)^6}\frac{2\pi^2}{16\pi}T\Bigg[\(1-\frac{4m^2}{M_\phi^2}\)^3 \frac{\pi M_{\phi}^8}{\Gamma_\phi }& K_{1}\(\frac{M_{\phi}}{T}\) +\frac{2^{16}3^2 5\, T^{11}}{M_{\phi}^4}\Theta\(\frac{M_{\phi}}{10} - T\)  \\& +  2^8 3\,T^7 \Theta\(T-\frac{M_{\phi}}{10}\)\Bigg],
\end{align}
where here and below the Heaviside functions are put in by hand, and chosen to cut off the power-laws in the appropriate places. 
For $T\sim m$, this result breaks down. However, further expansions in
the low $T/m$ regime are possible to model the result in this region.

Similarly, we can calculate the energy transfer rate. With the above
approximation for the amplitude, this result can also be found exactly
in terms of Meijer G-functions; however, the result is not
particularly illuminating and we omit it here. The Meijer G-functions
can be expanded in the limit of large $T/m$ and $T/M_{\phi}$ to give to a good
approximation
\begin{align}\label{eqn:gaugembenergyxfer}\nn
n_1 n_2 \langle \sigma v (E_1 +E_2)\rangle =  \frac{1}{128\Lambda_a^2\Lambda_b^2}\frac{1}{(2\pi)^6}\frac{2\pi^2}{16\pi}T\Bigg[& \(1-\frac{4m^2}{M_\phi^2}\)^3 \frac{\pi M_{\phi}^9}{\Gamma_\phi }K_{2}\(\frac{M_{\phi}}{T}\) +\frac{2^{18}3^3 5 \, T^{12}}{M_{\phi}^4}\Theta\(\frac{M_{\phi}}{10} - T\)\\ &  \qquad +  2^{11}3 \,T^8 \Theta\(T-\frac{M_{\phi}}{10}\)\Bigg].
\end{align}
This is an excellent approximation for $T > m$. An expansion is
possible in the low $T/m$ limit which can capture this limit extremely
accurately. We do not present it here.

Away from the Maxwell-Boltzmann limit, with Bose-Einstein statistics,
eq.\ \eqref{eqn:finresExfer} is difficult to evaluate. However,
numerically we find that away from the resonance, the full result
using Bose-Einstein statistics is almost indistinguishable from
Maxwell-Boltzmann statistics. eq.\ \eqref{eqn:gaugembxsection} and
\eqref{eqn:gaugembenergyxfer} are reasonable approximations for the
full energy transfer rates found by evaluating eq.\ 
\eqref{eqn:finresExfer}. These approximations can be improved by
correcting the asymptotic power law behavior. By comparing comparing
the result of eq.\  \eqref{eqn:gaugembenergyxfer} to the numerically
evaluated eq.\ \eqref{eqn:finresExfer}, we find that the $T >
M_{\phi}$ asymptote should be rescaled by a factor of approximately
$4/\pi$, giving
\begin{align}\label{eqn:gaugembenergyxferBE}\nn
n_2n_1 \langle \sigma v (E_1 +E_2)\rangle =\frac{1}{128\Lambda_a^2\Lambda_b^2}\frac{1}{(2\pi)^6} \frac{2\pi^2}{16\pi}T\Bigg[\(1-\frac{4m^2}{M_\phi^2}\)^3 \frac{\pi M_{\phi}^9}{\Gamma_\phi }& K_{2}\(\frac{M_{\phi}}{T}\) +\frac{2^{18}3^3 5 \, T^{12}}{M_{\phi}^4}\Theta\(\frac{M_{\phi}}{10} - T\) \\ & + \frac{4}{\pi} 2^{11}3 \,T^8 \Theta\(T-\frac{M_{\phi}}{10}\)\Bigg].
\end{align}
In figure  \ref{fig:gauge-axion} we show the results of these
approximations.

\subsection{Fermion-scalar boson scattering}
\label{sec:fermion-scalar}

\begin{figure}[h!]
\centering
\includegraphics[scale = 0.4]{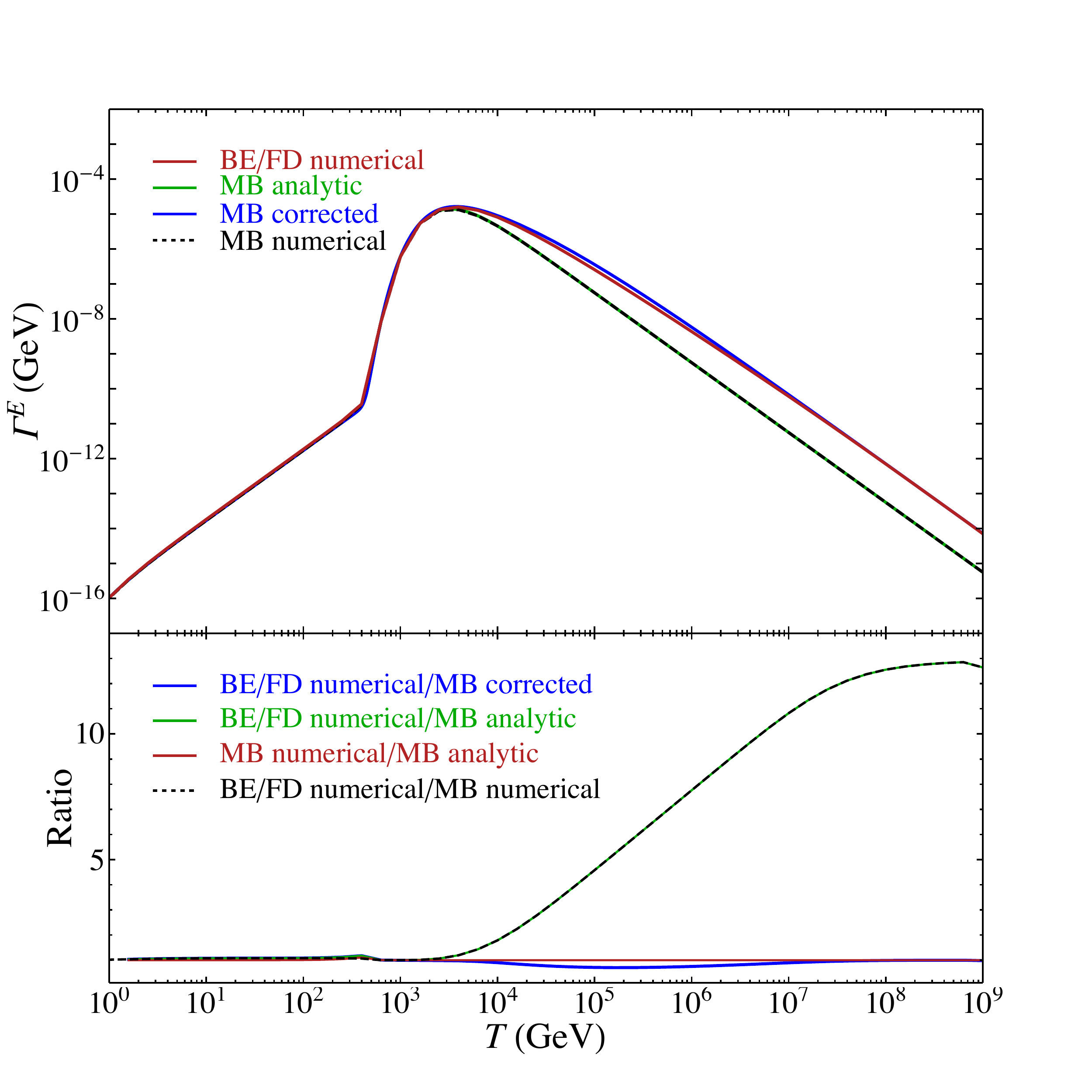}
\caption{Upper panel: We show the energy transfer rates for
  scalar-mediated fermion-to-scalar-field scattering. Parameters are
  chosen to be $m = m_S = m_{\psi}=1$ GeV, $M_{\phi} = 10^4$ GeV, $w =
  10^{-4}$. Shown are the rates for Fermi-Dirac/Bose-Einstein
  statistics (red), the numerically evaluated Maxwell-Boltzmann rate
  in black dashed and the analytic approximations at eq.\ 
  \eqref{eqn:FSMB} (green) and eq.\ \eqref{eqn:FSFDBEapprox}
  (blue). Lower panel: Black dashed shows the ratio of the
  Maxwell-Boltzmann energy transfer rate to the
  Fermi-Dirac/Bose-Einstein energy transfer rate. In red, we show the
  ratio of the approximation of eq.\ \eqref{eqn:FSMB} and the
  numerically evaluated energy transfer rate for Maxwell-Boltzmann
  statistics. In green we show the ratio of the approximation eq.\ 
  \eqref{eqn:FSMB} to the Fermi-Dirac/Bose-Einstein transfer
  rate. In blue we show the ratio of the approximation
eq.\  \eqref{eqn:FSFDBEapprox} to the Fermi-Dirac/Bose-Einstein transfer
  rate.}\label{fig:fermionscalar}
\end{figure}

Our model for fermion-scalar scattering is described in eqs.\ \ref{eq:hybridLagA1}--\ref{eq:hybridLagA3}.
The amplitude is well-approximated by
\begin{align}\nn
|\widebar{\mathcal{M}}(s)|^2\approx &2\mu^2y^2 \(1-\frac{4m^2}{s}\) \frac{s}{M_\phi^4}\Theta(M_\phi^2-s) +2\mu^2y^2 \frac{\pi}{\Gamma M_{\phi}}\(1-\frac{4m^2}{M_{\phi}^2}\)M_{\phi}^2\delta(s-M_{\phi}^2)\\
& +2\mu^2y^2 \(1-\frac{4m^2}{s}\) \frac{1}{s}  \Theta(s-M_\phi^2),
\end{align}
where $m = m_S = m_\psi$ is the mass of the scalar boson and the fermions, which we take to be equal.

In the limit where we treat all particles as classical, we can evaluate the
energy transfer rate. Again, we expand Meijer G-functions and keep
only the leading order power in $T$. The result is
\begin{align}\label{eqn:FSMB}
n_1 n_2 \langle \sigma v(E_1 + E_2)\rangle =&  \mu^2y^2  \frac{1}{(2\pi)^6} \frac{2\pi^2}{16\pi}T\Big[\frac{\pi  M_{\phi}^3}{\Gamma } \left(1-\frac{4 m^2}{M_{\phi}^2}\right)  K_2\left(\frac{M_{\phi}}{T}\right)\\ \nn&+ \Theta \left(\frac{M_{\phi}}{10}-T\right)\left(192  \frac{T^6}{M_{\phi}^4} - 128 \frac{m^2 T^2 }{M_{\phi}^4}- 16 \frac{T^2m^4}{M_{\phi}^4}\(4\gamma_E - 5 + 4\log\[\frac{m}{T}\]\) \right)\Big],
\end{align}
where $\gamma_E = 0.5772\ldots$ is the Euler-Mascheroni constant. The cross-section can also be found using the same approximations.

Again, the full integrals are difficult to integrate analytically. As
above, away from the resonance we can correct the Maxwell-Boltzmann
results, finding
\begin{align}\nn\label{eqn:FSFDBEapprox}
n_1 n_2 \langle \sigma v(E_1 + E_2)\rangle = &\mu^2y^2  \frac{1}{(2\pi)^6} \frac{2\pi^2}{16\pi}T\Big[\frac{\pi  M_{\phi}^3}{\Gamma } \left(1-\frac{4 m^2}{M_{\phi}^2}\right)  K_2\left(\frac{M_{\phi}}{T}\right)\(1+A e^{-B \sqrt{\frac{M}{T}}}\Theta(T - M_{\phi})\)\\ & \qquad +  \Theta \left(\frac{M_{\phi}}{10}-T\right)\left(192  \frac{T^6}{M_{\phi}^4} - 128 \frac{m^2 T^2 }{M_{\phi}^4}- 16 \frac{T^2m^4}{M_{\phi}^4}\(4\gamma_E - 5 + 4\log\[\frac{m}{T}\]\) \right)\Big],
\end{align}
where $A \approx 12$ and $B\approx 5/2$. The results of these
approximations are shown in figure  \ref{fig:fermionscalar}.  In this
example, the Maxwell-Boltzmann limit is functionally slightly
different from the full result as one moves away from the
resonance. This is well accounted for by the factor of $\(1+2\pi
e^{-\sqrt{\frac{M}{T}}}\Theta(T - M_{\phi})\)$; however, the
given values of the coefficients $A$ and $B$ are accurate only at the $\sim
\mathcal{O}(1)$ level as the parameters $M_{\phi}$ and $\Gamma_\phi$
are varied. Of all the cases considered, this case was the most
difficult to handle numerically, and the above presented result is
accurate only at the $10\%$ level.

\subsection{Fermion-gauge boson scattering}
\label{sec:fermion-gauge}

\begin{figure}[h!]
\centering
\includegraphics[scale = 0.4]{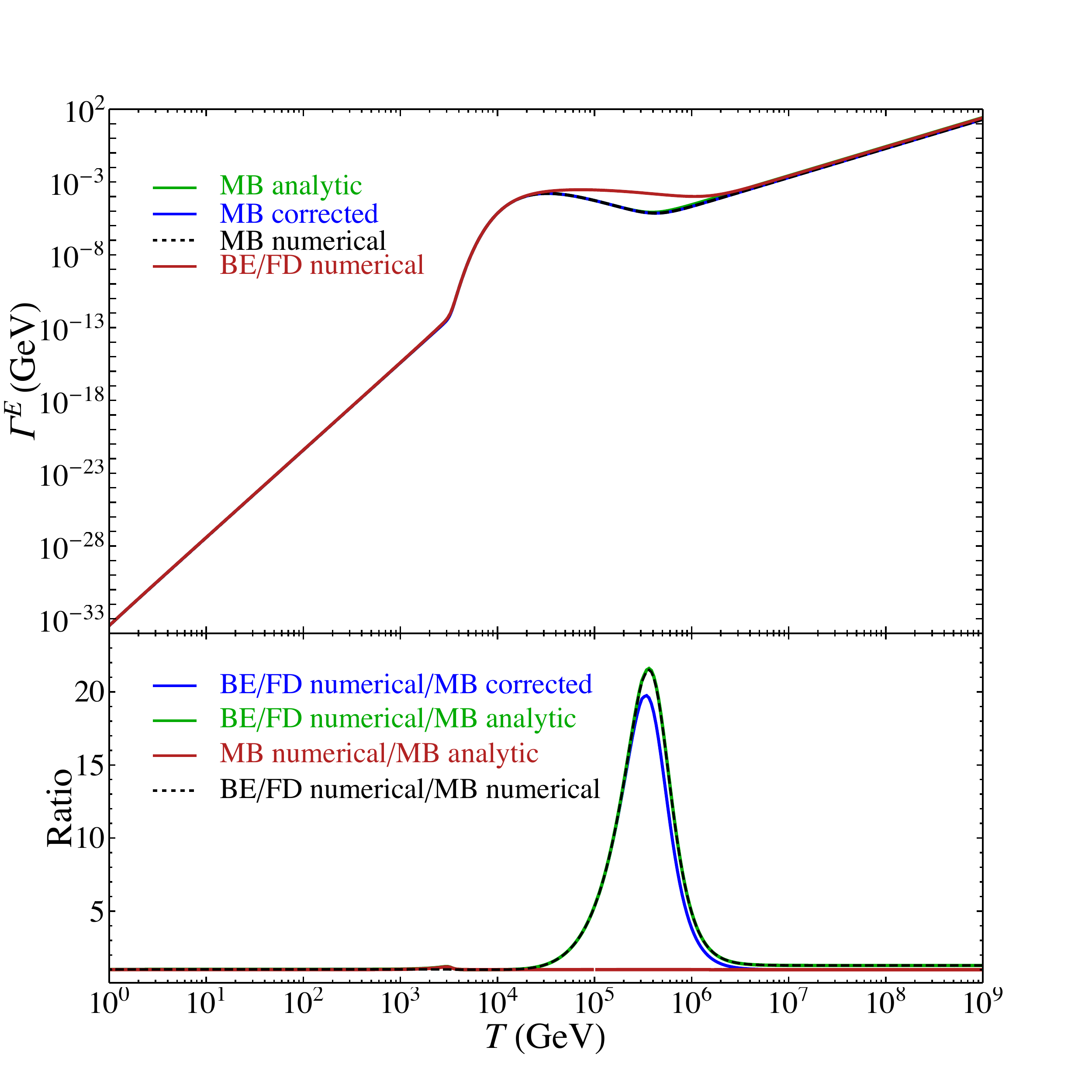}
\caption{Upper panel: We show the energy transfer rate for
  axion-mediated fermion-to-gauge-field scattering. Parameters are
  chosen to be $m = m_\psi = m_\gamma= 1$ GeV, $M_{\phi} = 10^5$ GeV,
  $w = 10^{-4}$. Shown are the rate for Fermi-Dirac statistics (red),
  the numerically evaluated Maxwell-Boltzmann rate in black dashed and
  the analytic approximations of eq.\ \eqref{eqn:fermion-gaugexferMB}
  (green) and eq.\ \eqref{eqn:fermion-gaugexferQ} (blue). Lower
  panel: Black dashed shows the ratio of the Maxwell-Boltzmann energy
  transfer rate to the Fermi-Dirac energy transfer rate. In red, we
  show the ratio of the approximation at eq.\ 
  \eqref{eqn:fermion-gaugexferMB} and the numerically evaluated energy
  transfer rate for Maxwell-Boltzmann statistics. In green we show the
  ratio of the approximation at eq.\ \eqref{eqn:fermion-gaugexferMB} to the
  Fermi-Dirac transfer rate. In blue we show the ratio of the
  approximation eq.\ \eqref{eqn:fermion-gaugexferQ} to the Fermi-Dirac
  transfer rate.}\label{fig:gauge-fermion}
\end{figure}

Our model for axion-mediated gauge boson-fermion interactions is
described in eqs.\ \eqref{eq:hybridLagB1}--\eqref{eq:hybridLagB2}.
Analogous to the above cases, this matrix element is well-approximated
by
\begin{align}\nn
\left|\widebar{\mathcal{M}}(s)\right|^2 \approx &\frac{m_{\psi}^2}{16\Lambda_a^2\Lambda_b^2} \(1-\frac{4m^2}{s}\) \frac{s^3}{M_\phi^4}\Theta(M_\phi^2-s) + \frac{m_{\psi}^2}{16\Lambda_a^2\Lambda_b^2}\frac{\pi}{\Gamma M_{\phi}}\(1-\frac{4m^2}{M_{\phi}^2}\)M_{\phi}^6\delta(s-M_{\phi}^2)\\
& +\frac{m_{\psi}^2}{16\Lambda_a^2\Lambda_b^2}\(1-\frac{4m^2}{s}\)^2  \Theta(s-M_\phi^2),
\end{align}
where $m = m_\gamma = m_\psi$ is the mass of the fermions and gauge bosons, which we are taking to be equal. 

In the limit where we ignore the details of quantum statistics, we can
integrate the energy transfer rate to obtain
\begin{align}\label{eqn:fermion-gaugexferMB}
n_1 n_2 \langle \sigma v(E_1 + E_2)\rangle = &\frac{m_{\psi}^2}{8\Lambda_a^2\Lambda_b^2} \frac{1}{(2\pi)^6} \frac{2\pi^2}{16\pi}T\Big[2^6 3 T^6 \Theta (T-M_{\phi}) +\frac{\pi  M_{\phi}^7}{\Gamma } \left(1-\frac{4 m^2}{M_{\phi}^2}\right)  K_2\left(\frac{M_{\phi}}{T}\right)\\ \nn& +  \Theta \left(\frac{M_{\phi}}{10}-T\right) \left( \frac{2^8 m^{10}}{15 M_{\phi}^4}-\frac{2^7 m^8 T^2}{3M_{\phi}^4}+\frac{2^{10} 3 m^4 T^6}{M_{\phi}^4}-\frac{2^{14} 3 m^2 T^8}{M_{\phi}^4}+\frac{2^{13} 3^2 5 T^{10}}{M_{\phi}^4}\right)\Big],
\end{align}
where the analytic result has been expanded in the limits $T \gg m$ and $T \gg M_{\phi}$. The cross-section can also be found using the same approximation.

Away from the Maxwell-Boltzmann limit, analytic evaluation of the full
energy transfer integral with full quantum statistics is
difficult. However, as described above, we can compare the full result
found by numerically evaluating eq.\ \eqref{eqn:finresExfer} with our
Maxwell-Boltzmann approximation in eq.\ 
\eqref{eqn:fermion-gaugexferMB} and correct the asymptotics to find
\begin{align}\label{eqn:fermion-gaugexferQ}
n_1 n_2 \langle \sigma v(E_1 + E_2)\rangle = &\frac{m_{\psi}^2}{8\Lambda_a^2\Lambda_b^2}\frac{1}{(2\pi)^6} \frac{2\pi^2}{16\pi}T\Big[A \,2^6 3 T^6 \Theta \(T-\frac{M_{\phi}}{10}\) +\frac{\pi  M_{\phi}^7}{\Gamma } \left(1-\frac{4 m^2}{M_{\phi}^2}\right)  K_2\left(\frac{M_{\phi}}{T}\right)\\\nn & +  \Theta \left(\frac{M_{\phi}}{10}-T\right) \left( \frac{2^8 m^{10}}{15 M_{\phi}^4}-\frac{2^7 m^8 T^2}{3M_{\phi}^4}+\frac{2^{10} 3 m^4 T^6}{M_{\phi}^4}-\frac{2^{14} 3 m^2 T^8}{M_{\phi}^4}+\frac{2^{13} 3^2 5 T^{10}}{M_{\phi}^4}\right)\Big]
\end{align}
where $A \approx 1.29$.  The results of these approximations are shown in figure  \ref{fig:gauge-fermion}. We note that away from the $T = M_{\phi}$, this is an excellent approximation.

\section{Preheating}\label{sec:preheating}

In this appendix we briefly review the physics of preheating and
sketch out the regions of parameter space in each of our simple
reheating models where we expect that preheating may be important. The
theory of preheating is a well-developed field; see, for example,
refs.\ \cite{Bassett:2005xm, Allahverdi:2010xz, Amin:2014eta} for
recent reviews.

\subsection{Preheating in brief}

The details of the reheating phase that immediately follows inflation
are model-dependent.  In particular, the overall description of
particle production is dependent on both details of the {\em
  inflationary} model, such as the scale at the end of inflation
$V(\phi_i)$ and the structure of the potential probed by the field
oscillations, as well as on the details of the {\em particle} model
describing the couplings of the inflaton to other degrees of freedom.
In this section we briefly present the general conditions for the
universe to undergo a period of strong, non-perturbative particle
production, dubbed \emph{preheating}, sourced by the time-dependent
inflaton background immediately following inflation, and discuss how
to understand these general conditions in the parameter space
presented in section \ref{sec:reheatT}.

Preheating can lead to rapid particle production through the
parametric resonance of momentum modes in the daughter fields which
couple to the inflaton. The oscillating inflaton background causes non-adiabatic changes in the effective frequency of these daughter fields which can result in their copious production. This non-perturbative particle production can
be, but is not always, highly efficient at transferring energy from
the inflaton condensate into radiation, thereby resulting in a much
higher reheating temperature than that obtained from estimates based on
perturbative decay.  By contrast, when preheating is incomplete (in
the sense that it cannot dissipate an $\mathcal{O}(1)$ fraction of the
initial energy density from the condensate), it will be followed by a
phase of perturbative reheating, and the ultimate impact of preheating
on $\Trh$ will typically be small.

In our study, we have assumed that at the beginning of the reheating
phase, the energy density of the universe is overwhelmingly dominated
by a single scalar field with a quadratic potential. When the Hubble
rate of the universe drops below the mass of the scalar field,
$M_{\phi}$, it begins to oscillate approximately sinusoidally with a
decaying envelope.  While the expansion of the universe is extremely
important for determining whether preheating is an effective means of
transferring energy out of the inflaton condensate, the essential
physics of the preheating process can be understood in flat space. In
this limit, daughter fields $\chi$ coupled to the oscillating inflaton
can typically have their equations of motion recast as the Mathieu
equation,
\begin{align}
\chi_k'' + (A_k - 2 q \cos(2z))\chi_k = 0.
\end{align}
Here the derivative acts over $z = M_{\phi}t + \delta$, where $\delta$ is a phase, and $A_k$
and $q$ depend on the parameters of the theory, such as the initial
background field amplitudes and the couplings and masses of the
external field(s) $\chi$.  The Mathieu equation can be solved exactly.
The standard arguments used in the analysis of this equation are
reminiscent of those used to solve the Schrodinger wave equation in a
periodic potential, with solutions given by Bloch waves,
\begin{align}
\chi_k(z) = e^{m_k z}P(k, z)
\end{align}
where $P$ is periodic in $z$ with period $\pi$. The quantity $m_k$ is
called the Mathieu exponent, and its properties determine the
stability of the solutions.

For growing solutions, the real part of the Mathieu exponent $\Re[m_k] = \mu_k$ is always non-negative. If $\mu_k(q) = 0$ then $|\chi_k|$ is
stable while if $\mu_k(q) > 0$, then the field amplitude
$|\chi_k|$ grows exponentially. Contour plots of $\mu_k$ are usually
presented as a function of the parameters $A$ and $q$, which together
control the width of the unstable bands and their strength
$\mu_k$. In general it is useful to distinguish two regimes:
\begin{itemize}
\item The {\it narrow resonance} regime, where $q < 1$. In this regime, $\mu_k
  \ll 1$, and the growth of particle occupation numbers can be
  moderate compared to the expansion of spacetime.
\item The {\it broad resonance} regime, where $q>1$. In this regime $\mu_k \gtrsim
  \mathcal{O}(1)$, and particle production is explosive.
\end{itemize}
The occupation numbers in the unstable bands grow exponentially as
\begin{align}
n_k \sim \exp(2\mu_k z) = \exp(2\mu_k M_\phi t).
\end{align}
In an expanding spacetime, the redshifting of momentum modes will
modify this picture, most importantly by shutting off the parametric
resonance. Two necessary conditions for preheating to be successful,
i.e., for an $\mathcal{O}(1)$ fraction of the inflaton energy density
to be dissipated through preheating, are:
\begin{enumerate}

\item The rate at which modes are amplified must proceed faster than
  the decay rate of the inflaton, and faster than the rate at which the inflaton oscillations are damped
\begin{align}\label{eqn:weakpreheat}
\frac{\Gamma}{M_{\phi}},\frac{H}{M_\phi}  < q .
\end{align}

\item The rate at which preheating proceeds must be faster than the
  rate at which the modes are redshifted out of the resonance
  band. The time that the mode remains in the resonance band depends
  on the equation of state of the matter; however, in the narrow
  resonance regime, it can be estimated as
  \cite{Kofman:1994rk}
\begin{align}
\Delta t \sim q H^{-1}.
\end{align}
Requiring that the parametric resonance rate exceeds this, one finds
the condition
\begin{align}\label{eqn:instabcond}
q^2 M_{\phi} \gtrsim H.
\end{align}

\end{enumerate}
Typically in the narrow resonance region ($q < 1$) eq.\ 
\eqref{eqn:instabcond} is a stronger condition than is eq.\ 
\eqref{eqn:weakpreheat}. Both conditions are satisfied in the broad
resonance region $q>1$.  Unless preheating begins with $q \gg 1$, and
ends near $q \sim 1$, it is unlikely to be completely effective at
reheating the universe, and it is expected that the final reheating
temperature will be determined by perturbative decays
\cite{Allahverdi:2010xz}.  However, preheating will alter the details
of the distribution functions of fields that are coupled to the
inflaton and can affect the timescale for the radiation bath to attain
internal thermal equilibrium.

The parameter regions relevant for preheating depend on the specific
model in question, so we will consider each case in turn.

\subsection{Trilinear scalar couplings}
\label{sec:preheating_scalars}

The works of refs.\ \cite{Shtanov:1994ce, Dufaux:2006ee} consider the
case of reheating through the trilinear interaction
\begin{align}
\mathcal{L} \supset   \frac{\mu}{2}\phi S^2+\ldots,
\end{align}
In this case, the effective frequency of the $S$ particles in the
oscillating $\phi$ background is
\begin{align}
\omega_k^2 = k^2+m_{S}^2 + \mu \phi(t) \approx k^2 + \mu \phi(t).
\end{align}
In this case, modes with $k^2+m_{S}^2 < \mu\Phi$ are exponentially
enhanced, where here and throughout $\Phi$ is the initial inflaton
amplitude.
For this model, we can identify the Mathieu parameters
\begin{align}
A_k = \frac{4k^2}{M_{\phi}^2}, \quad q = \frac{2\mu \Phi}{M_{\phi}^2}.
\end{align}
Since for $A_k < 2q$ the frequency of modes with momentum $k$ can
become negative during part of the oscillation period, there is a
tachyonic instability in this model. The regime in which this
instability is important is known as `tachyonic resonance.' When $q
\gg 1$, many dynamical modes are tachyonic for part of the
oscillation, and the resulting period of preheating is extremely
efficient and reheating can occur within a few oscillations of the
inflaton. However, provided $q < 1$, the width of the stability and
instability bands are comparable, and tachyonic resonance becomes
indistinguishable from narrow parametric resonance\footnote{As defined
  in \cite{Kofman:1994rk}, narrow parametric resonance applies when
  $A_k>2q$, as can happen for large $k$ or if the scalars $S$ had
  non-negligible masses.}
\cite{Dufaux:2006ee}. 

As the criteria for efficient preheating depend on both $\mu$ and
$\Phi$ for a fixed value of $M_\phi$, specifying $M_\phi$ and the
perturbative value of $\Trh$ is not sufficient to determine whether or
not preheating is important in a given model: the cosmological
history, i.e., the field amplitude $\Phi$ at the onset of reheating,
must also be specified.  As discussed in section \ref{sec:scalartherm},
only a bounded range for $\Phi$ is consistent with the minimal model
of reheating that we adopt here, which allows us to broadly
distinguish regions in the parameter space $(M_\phi,\,\Trh)$ where
preheating is, may be, or is not relevant.

\subsection{Axion couplings to gauge fields}
\label{sec:preheating_axion}

In this work we have considered the coupling of an axion to gauge
fields via the dimension-5 operator
\begin{align}
\mathcal{L} = \frac{\phi}{4 \Lambda} F_{\mu\nu}\tilde{F}^{\mu\nu},
\end{align}
where $\Lambda$ is a mass scale associated with the axion. This case
is unique, because it involves a derivative coupling to an irrelevant
operator.

\subsubsection*{Preheating constraints}

This case is treated in refs.~\cite{ArmendarizPicon:2007iv, Adshead:2015pva}.
The equation of motion for the gauge modes, $B^{\pm}_k$, in flat space is given by
\begin{align}
\ddot{B}^{\pm}_{k} + k\(k\mp \frac{\dot{\phi}}{\Lambda}\){B}^{\pm}_{k}  = 0
\end{align}
where $\pm$ denotes the two possible helicity states of the gauge
field, and the derivatives act over time $t$.  Defining $z = M_\phi t$, and writing $\dot{\phi} = \mp M_{\phi}
\Phi \cos(2 z)$, we can again identify the Mathieu parameters:
\begin{align}
A_k = \frac{4k^2}{M_{\phi}^2}, \quad q = 2\frac{k}{M_{\phi}}\frac{\Phi}{\Lambda}.
\end{align}
The different signs for the two different helicities have been
absorbed into the choice of phase. 
Note that it may appear that one can get large $q$ here, and thus
efficient preheating, by looking at larger wavenumbers. However, as
one increases $k$, $A$ also increases and pushes one into the stable
regions of the Mathieu chart. Parametric resonance in these models is
most effective for the \emph{lowest} momentum modes possible, and thus
we will use the value of $q$ for the lowest dynamical mode, $k \sim a
H\sim M_\phi \Phi/\Mp$. For this mode, we have $A_k < 2q$, which is
similar to the tachyonic resonance case considered for scalars
above. The narrow resonance condition becomes
\begin{align}
q < 1 \quad \Rightarrow \quad 2 \frac{\Phi^2}{\Mp^2} < \frac{\Lambda}{\Mp} ,
\end{align}
while the condition that narrow resonance is inefficient is
\begin{align}
q^2 M_{\phi} < H \quad \Rightarrow \quad 2 \frac{\Phi^3}{M_{\rm Pl}^3} < \frac{\Lambda^2}{M_{\rm Pl}^2}.
\end{align}
For large field inflation, $\Phi \sim M_{\rm Pl}$, notice
that these conditions roughly coincide.

\subsubsection*{Inflationary constraints}

While in the present work we have been agnostic about whether the
field that drives reheating is the same field that is responsible for
inflation, it is worth pointing out that in the case where the axion
does drive inflation, the coupling to gauge fields can lead to the
exponential production of gauge bosons during the inflationary phase
itself \cite{Carroll:1991zs, Garretson:1992vt}. These gauge bosons can
scatter off the inflaton condensate and generate fluctuations in the
axion field. The axion fluctuations generated this way are strongly
non-Gaussian, and can lead to unacceptable levels of non-Gaussianity
in the CMB \cite{Barnaby:2011vw, Barnaby:2011qe} and primordial black
holes at the end of inflation \cite{Linde:2012bt} if the couplings are
too large.

The conditions on the inflationary production of gauge quanta are
quoted in terms of a parameter $\xi$, defined as
\begin{align}
\xi = \frac{1}{2}\frac{\dot{\phi}}{H}\frac{1}{\Lambda} \equiv \sqrt{\frac{\epsilon_H}{2}}\frac{M_{\rm Pl}}{\Lambda}.
\end{align}
The bounds from non-Gaussianity in the CMB \cite{Ade:2015ava} are $\xi
< 3.3$ during the observable $e$-folds of evolution. We can use the
constraint on the tensor-to-scalar ratio $r < 0.07 -0.09$
\cite{Array:2015xqh}, and the slow-roll condition
$ 
r  = 16\epsilon_H
$ 
to place a bound on the scale $\Lambda$,
\begin{align}
\frac{M_{\rm Pl}}{\Lambda} < 3.3\sqrt{\frac{32}{0.07}} \approx 70,\quad
\text{  or  }\quad
\frac{M_{\rm Pl}}{70} < \Lambda.
\end{align}
Again, this bound is only applicable if the axion in question is the
field that drives the period of slow-roll inflation constrained by the
CMB.

\subsection{Fermion Preheating}
\label{sec:preheating_fermions}

The preheating of fermions via a Yukawa coupling of the form
\begin{align}
\mathcal{L} = y_i \phi \bar{\psi}_i \psi_i
\end{align}
was first studied in detail by Kofman and Greene
\cite{Greene:2000ew}. The efficiency of fermion preheating is limited
by the fact that fermions obey the Pauli exclusion principle, which
prevents them piling up in one state. However, fermion preheating can
be somewhat effective at populating fermion states.

For the Yukawa theory, the analogue of the $q$ parameter is given by
\begin{align}
 q = \frac{y^2 \Phi^2}{M_\phi^2} .
\end{align}
Preheating fills up fermion states up to a maximum wave number of \cite{Greene:2000ew}
\begin{align}
k_{\rm max} \sim a^{1/4}q^{1/4} M_{\phi},
\end{align}
which grows like $a^{1/4}$. While in the bosonic case, taking $q>1 $
leads to efficient `broad resonance' preheating, for fermions the effect of a large $q$ parameter is to excite
modes up to a higher wavenumber.

In order to see if fermion preheating can significantly alter our
conclusions, we examine the amount of energy that is transferred
non-perturbatively from the inflaton condensate to fermions. The
energy density in the preheated fermions can be estimated as
\begin{align}
\rho_{\psi} = \int \frac{d^3 k}{(2\pi)^3} n_k E_k.
\end{align}
Taking $E_k \sim k/a$, and $n_k = 1/2$ up to $k_{\rm max}$, we have
\begin{align}
\rho_{\psi} = \frac{1}{(2\pi)^2} \frac{1}{a^4}\int_0^{k_{\rm max}} k^3 dk  =  \frac{1}{4(2\pi)^2a^3}q M_\phi^4 =  \frac{1}{4(2\pi)^2}y^2 \frac{\Phi^2 M_\phi^2}{a^3}.
\end{align}
In the absence of their perturbative decay, the preheated fermions
remain a fixed (small) fraction of the total energy density.  For
$m_{\psi} \ll M_{\phi}$, the zero-temperature inflaton width into
fermions is given by
\begin{align}
\frac{\Gamma}{M_{\phi}} \simeq \frac{y^2}{8\pi },
\end{align}
so we can express the above as
\begin{align}
\rho_{\psi} = \frac{\Gamma}{M_{\phi}}\frac{1}{2\pi}\rho_{\phi}.
\end{align}
Thus for perturbative couplings the impact of preheated fermions on
our calculations is negligible.

  \bibliographystyle{JHEP}
  \bibliography{dm_reheating}

\end{document}